\documentclass
[superscriptaddress,secnumarabic,amssymb,amsmath,nobibnotes,aps,prd,showkeys,showpacs,nofootinbib,onecolumn]{revtex4}%
\usepackage{graphicx}
\usepackage{epsf}
\usepackage{bm}
\usepackage{amsmath}
\usepackage{amsfonts}
\usepackage{amssymb}
\usepackage{epstopdf}%
\providecommand{\U}[1]{\protect\rule{.1in}{.1in}}

\newcommand{\be}{\begin{equation}}
\newcommand{\ee}{\end{equation}}

\newcommand{\mincir}{\raise
-3.truept\hbox{\rlap{\hbox{$\sim$}}\raise4.truept\hbox{$<$}\ }}
\newcommand{\magcir}{\raise
-3.truept\hbox{\rlap{\hbox{$\sim$}}\raise4.truept\hbox{$>$}\ }}

\begin{document}
\title{Observational Constraints on New Exact Inflationary Scalar-field Solutions }
\author{John D. Barrow}
\email{jdb34@hermes.cam.ac.uk}
\affiliation{DAMTP, Centre for Mathematical Sciences, University of Cambridge, Wilberforce
Rd., Cambridge CB3 0WA, UK}
\author{Andronikos Paliathanasis}
\email{anpaliat@phys.uoa.gr}
\affiliation{Instituto de Ciencias F\'{\i}sicas y Matem\'{a}ticas, Universidad Austral de
Chile, Valdivia, Chile}

\begin{abstract}
An algorithm is used to generate new solutions of the scalar field equations
in homogeneous and isotropic universes. Solutions can be found for pure scalar
fields with various potentials in the absence and presence of spatial
curvature and other perfect fluids. A series of generalisations of the
Chaplygin gas and bulk viscous cosmological solutions for inflationary
universes are found. Furthermore other closed-form solutions which provide
inflationary universes are presented. We also show how the Hubble slow-roll
parameters can be calculated using the solution algorithm and we compare these
inflationary solutions with the observational data provided by the Planck 2015
collaboration in order to constraint and rule out some of these models.

\end{abstract}
\keywords{Cosmology; Scalar field; Inflation; Chaplygin gases}
\pacs{98.80.-k, 95.35.+d, 95.36.+x}
\maketitle
\date{\today}

\section{Introduction}

Recent cosmological data indicate that the universe has undergone two
acceleration phases: an early acceleration phase called 'inflation', prior to
a radiation-dominated era, and a more recent era of accelerated expansion
which appears to continue today
\cite{Teg,Kowal,Komatsu,suzuki11,Ade15,planck2013,planck2015}. The
gravitationally repulsive stress that is responsible for the current
acceleration of the universe is called 'dark energy' and must possess
sufficient negative pressure to exert gravitational repulsion. Its specific
identity is still unknown and it may result from a modification of general
relativity when gravity is very weak or the presence of a specific unknown
matter field.

Whilst a range of \textquotedblleft exotic\textquotedblright\ fluids and
modifications of the gravitational action can provide cosmological
acceleration, scalar fields are the simplest candidates to explain the
acceleration phases of the universe. Moreover, scalar fields also have various
applications in the inflationary phase of the universe, for instance in
driving chaotic inflation \cite{lid02}. While the same scalar field might
explain both of the periods of accelerated expansion, no convincing
cosmological model has been found with this as a natural feature. In a scalar
field cosmology the field equations are of second-order where the scalar field
is introduced an extra degree of freedom, with a corresponding conservation
equation. These equations display unexpected complexity. Simple power-law
potentials for the scalar field can create finite-time singularities during
inflation \cite{sing} and lead to chaotic dynamics \cite{page}, or singularity
avoidance \cite{BMatz} if the universe is closed.

Very few exact scalar-field solutions in a
Friedmann-Lema\^{\i}tre-Robertson-Walker spacetime (FLRW) with spatial
curvature are known \cite{hall,Ea}. In a spatially-flat FLRW spacetime
closed-form solutions with or without sources for different scalar field
potentials, or scalar fields which mimic other fluids, such as the Chaplygin
gas, can be found in \cite{jdbchg,
jdbmin,jdbnew,muslinov,ellis,barrow1,Mendez,barrowhyp,chimento,newref2,rubano,sahni,Kahya,basil,tscqg,palprd2015,pan}
while some other classes of integrable scalar-field models are also given
\cite{suzuki,fre,palia1,newref1}. Some solutions for three-dimensional FLRW
spacetimes are given in \cite{bblanc,william,shaw}. However, scalar-field
cosmology is conformally equivalent to other scalar-tensor theories, like
Brans-Dicke or $f\left(  R\right)  $-gravity. \cite{bcot, maeda}. Hence,
closed-form solutions of the conformally equivalent theories (see
\cite{sol3,sol4} and references therein) can be used to construct closed-form
solutions or to find new integrable systems for the non-minimally coupled
scalar-field model.

Recently, with the use of nonlocal conservation laws in \cite{dimakis}, the
general analytical solution has been expressed for an arbitrary scalar field
with an arbitrary number of independent perfect fluids possessing constant
equation of state parameters in spatially flat or nonflat FLRW universes.
These general results are applied in this paper to derive precise forms of the
scalar field potential for various simple time-dependent forms for the
expansion scale factor, or for particular equation of state parameters for the
scalar field. Finally, the Hubble slow-roll parameters are studied for these
closed-form solutions so that we can compare the inflationary parameters with
the observable constraints provided by the Planck 2015 observations
\cite{planck2015}.

\ The plan of this paper is as follows. In section \ref{field}, the basic
properties and definitions of scalar-field models are introduced. The
cosmological metric we consider is the four-dimensional FLRW spacetime, while
the gravitational action is that of general relativity with a minimally
coupled scalar field. We review previous results in the literature and present
the general analytical solution for the cosmological field equations for
arbitrary scalar-field potential. Exact closed-form solutions obtained by
using these general results are presented in sections \ref{exactS1} and
\ref{exactS2}. Specific closed-form solutions are derived for spatially-flat
FLRW universes when only one scalar field and a perfect fluid with constant
equation of state parameter are present. For specific values of the barotropic
parameter for the matter source, these results give closed-form solutions in
the case of a nonflat FLRW universe. In section \ref{hsrpar}, we derive the
Hubble slow-roll parameters for our models and compare them with that of the
Planck 2015 data to isolate observationally allowed parameter ranges. Finally,
in section \ref{conc}, we discuss our results and draw conclusions.

\section{ \ Scalar-field cosmology}

\label{field}

We consider the gravitational action to be
\begin{equation}
S=S_{EH}+S_{\phi}+S_{m}, \label{SF.01}%
\end{equation}
in which $S_{EH}=\int dx^{4}\sqrt{-g}R~$is the Einstein-Hilbert action, $R$ is
the Ricci Scalar of the underlying spacetime geometry with metric tensor
$g_{\mu\nu}$, $S_{m}=\int dx^{4}\sqrt{-g}L_{m}$ is the matter action, and
$S_{\phi}$ is the action for the scalar field, with
\begin{equation}
S_{\phi}=\int dx^{4}\sqrt{-g}\left[  -\frac{1}{2}g^{\mu\nu}\phi_{;\mu}%
\phi_{;\nu}+V(\phi)\right]  , \label{SF.02}%
\end{equation}
where $V(\phi)$ is the self-interaction potential of the scalar field $\phi
.$Variation of $S$ with respect to $g_{\mu\nu}$ gives the Einstein equations,%
\begin{equation}
R_{\mu\nu}-\frac{1}{2}g_{\mu\nu}R=T_{\mu\nu}^{\left(  \phi\right)  }+T_{\mu
\nu}^{\left(  m\right)  }, \label{sfa.01}%
\end{equation}
where\ $R_{\mu\nu}$ is the Ricci tensor,\ $T_{\mu\nu}$ is the energy-momentum
tensor of baryonic\ matter and radiation, and $T_{\mu\nu}(\phi)$ is the
energy-momentum tensor associated with $\phi$. Furthermore, variation with
respect to $\phi$ gives%
\begin{equation}
-g^{\mu\nu}\phi_{;\mu\nu}+V_{,\phi}=0, \label{sfa.02}%
\end{equation}
where we have considered that $\frac{\partial S_{m}}{\partial\phi}=0$, so
there is no interaction between the matter source, $S_{m}$, and the scalar
field, $\phi$, in the action integral (\ref{SF.01})

Using the Bianchi identity for (\ref{sfa.01}) we have that $T_{~~~~~~~\ ;\nu
}^{\left(  \phi\right)  \mu\nu}+T_{~~~~~~~\ ;\nu}^{\left(  m\right)  \mu\nu
}=0,~$which gives%
\begin{equation}
T_{~~~~~~~\ ;\nu}^{\left(  \phi\right)  \mu\nu}=0~,~T_{~~~~~~~\ ;\nu}^{\left(
m\right)  \mu\nu}=0. \label{sfa.03}%
\end{equation}
These are the equations of motion for the matter sources $S_{m}$ and the field
$\phi$. It can be seen that (\ref{sfa.03}) is just equation (\ref{sfa.02}).

By assuming that the universe is spatial isotropic and homogeneous we select
the four-dimensional spacetime to be that of FLRW
\begin{equation}
ds^{2}=-dt^{2}+a^{2}(t)\frac{(dx^{2}+dy^{2}+dz^{2}).}{(1+\frac{\mathit{K}}%
{4}\mathbf{x}^{2})^{2}}. \label{SF.1}%
\end{equation}

Furthermore, we assume that $\phi$ inherits the symmetries of the metric
(\ref{SF.1}). Therefore $\phi(t)$ and consequently $\phi_{;\nu}=\dot{\phi
}\delta_{\nu}^{0}$, where $\dot{\phi}=\frac{d\phi}{dt}$. \ Consider the
comoving observer $u_{\mu}=\delta_{t}^{\mu}$, $\left(  u^{\mu}u_{\mu
}=-1\right)  $. In the 1+3 decomposition the energy-momentum tensor becomes%
\begin{equation}
T_{\mu\nu}^{\left(  \phi\right)  }=\left(  \rho_{\phi}+P_{\phi}\right)
u_{\mu}u_{\nu}+P_{\phi}g_{\mu\nu},
\end{equation}%
\begin{equation}
T_{\mu\nu}^{\left(  m\right)  }=\left(  \rho_{m}+P_{m}\right)  u_{\mu}u_{\nu
}+P_{m}g_{\mu\nu},
\end{equation}
where
\begin{equation}
\rho_{\phi}\equiv\frac{1}{2}\dot{\phi}^{2}+V(\phi)~,~P_{\phi}\equiv\frac{1}%
{2}\dot{\phi}^{2}-V(\phi) \label{SF.04}%
\end{equation}
are the energy density and the pressure of the scalar field and $\rho
_{m},p_{m}$ are the components that correspond to the matter source $S_{m}$
which we have assumed to be a perfect fluid. This follows also from
(\ref{SF.1}).

Therefore, the field equations (\ref{sfa.01}) for the line-element
(\ref{SF.1}) become
\begin{equation}
H^{2}=\frac{1}{3}\left(  \rho_{m}+\rho_{\phi}\right)  -\frac{\mathit{K}}%
{a^{2}} \label{SF.07}%
\end{equation}%
\begin{equation}
3H^{2}+2\dot{H}=-(P_{m}+P_{\phi})-\frac{\mathit{K}}{a^{2}}, \label{frie2}%
\end{equation}
where $H(t)\equiv\frac{\dot{a}}{a}$ is the Hubble function.

Equations (\ref{sfa.03}) become
\begin{equation}
\dot{\rho}_{m}+3H(\rho_{m}+P_{m})=0\, \label{SF.08}%
\end{equation}%
\begin{equation}
\dot{\rho}_{\phi}+3H(\rho_{\phi}+P_{\phi})=0\, \label{SF.09}%
\end{equation}
while the corresponding equation of state (EoS) parameters are given by
$w_{m}=P_{m}/\rho_{m}$ and
\begin{equation}
w_{\phi}=\frac{P_{\phi}}{\rho_{\phi}}=\frac{(\dot{\phi}^{2}/2)-V(\phi)}%
{(\dot{\phi}^{2}/2)+V(\phi)}%
\end{equation}
which means that $w_{\phi}<-\frac{1}{3}$ when $\dot{\phi}^{2}<V(\phi)$. On the
other hand, if the kinetic term of the scalar field is negligible with respect
to the potential energy, i.e., $\frac{\dot{\phi}^{2}}{2}\ll V(\phi)$, then the
equation of state parameter is $w_{\phi}\simeq-1$.

Substituting (\ref{SF.04}) into (\ref{SF.09}), we find equation (\ref{sfa.02})
which for the line element (\ref{SF.1}) takes the form{{%
\begin{equation}
\ddot{\phi}+3H\dot{\phi}+V_{,\phi}=0. \label{SF.10}%
\end{equation}
}}

The set of equations, (\ref{SF.07}), (\ref{frie2}) and (\ref{SF.10}), provide
us with the cosmological evolution, i.e. the scale factor $a\left(  t\right)
$, where a potential $V\left(  \phi\right)  $ and an equation of state
parameter $w_{m}$ have been defined.

There is a simple recipe \cite{jdbchg,jdbmin,jdbnew,jdbpp1,jdbpp2} for finding
exact solutions in the flat FLRW universes containing only the scalar field
($\rho_{m}=p_{m}=0=\mathit{K}$), where the defining equations simplify to%

\begin{align}
3H^{2}  &  =\frac{1}{2}\dot{\phi}^{2}+V(\phi)\label{SF.10a}\\
2\dot{H}  &  =-\dot{\phi}^{2} \label{SF.10b}%
\end{align}
The third equation (\ref{SF.10}) is a consequence of these equations. The
recipe is to pick a physically realistic function $\phi(t)$,~solve
(\ref{SF.10b}) to find $H(t)$, use $H(t)$ and $\phi(t)$ to find $V(t)$ from
(\ref{SF.10a}) and convert this to $V(\phi)$ using the initial $\phi(t).$This
completes the solution so long as the intermediate integrals can be performed
and appropriate positivity conditions hold (for example, $H>0$ and $V\geq0$).
However, when a perfect fluid or 3-curvature (which is just another perfect
fluid) this method is not efficient and we must look to a more systematic
version. To this method we now turn.

\subsection{General analytical solution \emph{ }}

\label{gensf}

In the line element (\ref{SF.1}) we use the comoving proper time, $t$, by
putting $dt=N\left(  \tau\right)  d\tau$. From the action integral
(\ref{SF.01}), we can now define
\begin{equation}
L\left(  N,a,\dot{a},\phi,\dot{\phi}\right)  =\frac{1}{N}\left(  -3a\dot
{a}^{2}+\frac{1}{2}a^{3}\dot{\phi}^{2}\right)  -Na^{3}V\left(  \phi\right)
-N\rho_{m0}a^{-3\left(  \gamma-1\right)  }+3N\mathit{K}a, \label{SF.11}%
\end{equation}
where for the matter source, $S_{m}$, we have put $w_{m}=\gamma-1$. Hence the
gravitational field equations follow from the Euler-Lagrange equations with
respect to the variables $\left\{  N,a,\phi\right\}  $. However, as it can be
seen, the field equations in the space of variables $\left\{  N,a,\phi
\right\}  $ form a singular dynamical system \ with constraint equation
$\frac{\partial L}{\partial N}=0$. \emph{ }Hence, using \cite{dimakis} with
the application of the results of \cite{tdimakis}, it has been shown that the
gravitational field equations which follow from (\ref{SF.11}) admit an
infinite number of (nonlocal) conservation laws. Specifically, every conformal
Killing vector of the minisuperspace $\left\{  a,\phi\right\}  $ provides a
conservation law and, as the minisuperspace has dimension two, the dimension
of the conformal algebra is infinite and consequently we have an infinite
number of conservation laws. Here, it is important to note that these
conservation laws are not necessarily in involution. For more details see
\cite{tdimakis}.

With the use of a specific (nonlocal) conservation law, in \cite{dimakis} it
was proved that the field equations form an integrable system and for a
specific lapse, $\omega$, such as $dt=e^{F\left(  \omega\right)  /2}d\omega$,
where $a\left(  \omega\right)  \equiv e^{\omega/6}$; that is, the line element
is now
\begin{equation}
ds^{2}=-e^{F\left(  \omega\right)  }d\omega^{2}+e^{\omega/3}\frac
{(dx^{2}+dy^{2}+dz^{2})}{(1+\frac{\mathit{K}}{4}\mathbf{x}^{2})^{2}},
\label{SF.12}%
\end{equation}
the solution is expressed in terms of the unknown function $F\left(
\omega\right)  $, which is directly related to the potential $V\left(
\phi\right)  $.

In the case of a spatially flat universe, $\mathit{K}=0$, and without matter
source,$~\rho_{m0}=0$, it has been found that\footnote{Where a prime, i.e.
$F^{\prime}$,\ denotes derivative with respect to $\omega$.}%
\begin{equation}
\phi(\omega)=\pm\frac{\sqrt{6}}{6}\int\!\!\sqrt{F^{\prime}(\omega)}d\omega,
\label{so.01}%
\end{equation}
where%
\begin{equation}
V(\omega)=\frac{1}{12}e^{-F(\omega)}\left(  1-F^{\prime}(\omega)\right)
\label{so.02}%
\end{equation}
and effective fluid components for the scalar field are%
\begin{equation}
\rho_{\phi}(\omega)=\frac{1}{12}e^{-F(\omega)}~,~P_{\phi}(\omega)=\frac{1}%
{12}e^{-F(\omega)}\left(  2F^{\prime}(\omega)-1\right)  . \label{so.03}%
\end{equation}

Furthermore, in the case of a spatially flat universe with a perfect fluid the
solution is generalised as follows%
\begin{equation}
\phi(\omega)=\pm\frac{\sqrt{6}}{6}\int\left[  \left(  F^{\prime}\left(
\omega\right)  -6\gamma\rho_{m0}e^{F-\frac{\gamma}{2}\omega}\right)  \right]
^{1/2}\!d\omega, \label{so.04}%
\end{equation}
where now%
\begin{equation}
V(\omega)=\frac{1}{12}e^{-F(\omega)}\left(  1-F^{\prime}(\omega)\right)
+\frac{\gamma}{2}\,\rho_{m0}\,e^{-\frac{\gamma}{2}\omega} \label{so.05}%
\end{equation}
and the fluid components become%
\begin{equation}
\rho_{\phi}=\frac{1}{12}e^{-F(\omega)}-\,\rho_{m0}\,e^{-\frac{\gamma}{2}%
\omega} \label{so.06}%
\end{equation}
\mbox{\rm and}
\begin{equation}
P_{\phi}=\frac{1}{12}e^{-F(\omega)}\left(  2F^{\prime}(\omega)-1\right)
-\left(  \gamma-1\right)  \rho_{m0}\,e^{-\frac{\gamma}{2}\omega}.
\label{so.07}%
\end{equation}

In the latter case, the total fluid stress, $T_{\mu\nu}=T_{\mu\nu}^{\left(
\phi\right)  }+T_{\mu\nu}^{\left(  m\right)  }$, can be described by a new
field, $\Phi$, which follows from (\ref{so.01})-(\ref{so.03}). Also in the
latter, if we assume that $\gamma=\frac{2}{3}$ (to mimic a curvature term in
the Friedmann equations) and $\rho_{m0}=-3\mathit{K}$, then the solution of
the scalar-field model in a nonflat FLRW spacetime is recovered.

The aim of this work is to derive specific closed-form solutions of the field
equations using these results by assuming special inflationary functions for
the scalar factor, or special equation of state parameters for the scalar
field, which consequently combine to define the scalar-field potential.

\section{Closed-form solutions: spatially-flat FLRW spacetime without matter
source}

\label{exactS1}

Consider the simplest scenario for a spatially flat FLRW spacetime containing
only a scalar field. If we assume that the scalar field has a constant
equation of state parameter, say $w_{\phi}=\left(  -1+\frac{2}{3q}\right)  $,
where $q$ is a constant, then from (\ref{so.03}) we find that%
\begin{equation}
F\left(  \omega\right)  =2\ln\left[  \frac{e^{\frac{\omega}{6q}}}{6q\left(
a_{0}\right)  ^{\frac{1}{q}}}\right]  , \label{sd.01}%
\end{equation}
hence we have:
\begin{equation}
\phi\left(  \omega\right)  =\frac{\sqrt{2}}{6}q^{-\frac{1}{2}}\omega
~,~V\left(  \omega\right)  =V_{0}e^{-\frac{1}{3q}\omega}. \label{sd.02}%
\end{equation}
Therefore, $V\left(  \phi\right)  =V_{0}e^{-\sqrt{\frac{2}{q}}\phi}$ $=\left(
a_{0}\right)  ^{\frac{2}{q}}q(3q-1)e^{-\sqrt{\frac{2}{q}}\phi}$ in which, if
we apply the transformation $d\omega\rightarrow dt~$to write (\ref{SF.12}) in
the form of (\ref{SF.1}), we find the well-known power-law solution (which we
can verify directly in (\ref{SF.10a})-(\ref{SF.10b}): \emph{ }%
\begin{equation}
a\left(  t\right)  =a_{0}t^{q}~,~\phi=\phi_{0}+\sqrt{2q}\ln\left(  t\right)  .
\label{sd.03}%
\end{equation}

However, this is only a particular solution of the exponential scalar field
potential problem \cite{jbexp}. The general solution can be found in
\cite{russo}, while some special solutions are given in \cite{pied,bb}.

We continue with the determination of the closed-form solution for some
specific equation of state parameters for the scalar field.

\subsection{Perfect fluid with cosmological constant}

Assume that the scalar field satisfies the simple equation of state parameter%
\begin{equation}
p_{\phi}=\left(  \gamma-1\right)  \rho_{\phi}-3\gamma\Lambda H_{0}^{2}.
\label{sd.04}%
\end{equation}
Then, it follows from (\ref{so.03}), that
\begin{equation}
2F^{\prime}+36\gamma\Lambda e^{F}-\gamma=0. \label{sd.05}%
\end{equation}
We observe that, for $\Lambda=0$, $F\left(  \omega\right)  $ is linear as
above. Hence, for nonzero $\Lambda$ and $\gamma\neq0,$ we find that%
\begin{equation}
F\left(  \omega\right)  =-\ln\left(  36\Omega_{m0}H_{0}^{2}e^{-\frac{\gamma
}{2}\omega}+36\Omega_{\Lambda}H_{0}^{2}\right)  , \label{sd.06}%
\end{equation}
where $36\Omega_{m0}H_{0}^{2}$ is the constant of integration~and
$\Omega_{\Lambda}=\frac{\Lambda}{3H_{0}^{2}}$. \ The Hubble function is
\begin{equation}
\frac{H^{2}\left(  \omega\right)  }{H_{0}^{2}}=\Omega_{m0}e^{-\frac{\gamma}%
{2}\omega}+\Omega_{\Lambda}, \label{sd.07}%
\end{equation}
which is equivalent to a cosmological model containing a perfect fluid and a
cosmological constant. We can see that for $\gamma=1$, $\Lambda CDM$-cosmology
is recovered.

Furthermore, using (\ref{sd.06}) we find%
\begin{equation}
\phi\left(  \omega\right)  =-\frac{2}{\sqrt{3\gamma}}%
\text{\textrm{arctanh\thinspace}}\left(  \sqrt{\frac{\Omega_{m0}%
+\Omega_{\Lambda}e^{\frac{\gamma}{2}\omega}}{\Omega_{m0}}}\right)
\label{sd.08}%
\end{equation}
and%
\begin{equation}
V\left(  \omega\right)  =\frac{3}{2}H_{0}^{2}e^{-\frac{\gamma}{2}\omega
}\left(  \Omega_{m}\left(  2-\gamma\right)  +2\Omega_{\Lambda}e^{\frac{\gamma
}{2}\omega}\right)  , \label{sd.09}%
\end{equation}
where the equation of state parameter is
\begin{equation}
w_{\phi}\left(  \omega\right)  =-1+\gamma\frac{\Omega_{m0}}{\Omega_{m0}%
+\Omega_{\Lambda}e^{\frac{\gamma}{2}\omega}}. \label{sd.10}%
\end{equation}

Finally, we find the potential
\begin{equation}
V\left(  \phi\right)  =3\Lambda H_{0}^{2}+\frac{3\left(  2-\gamma\right)  }%
{2}\Lambda H_{0}^{2}\cosh^{2}\left(  \frac{\sqrt{3\gamma}}{2}\phi\right)  ,
\label{sd.11}%
\end{equation}
from which we observe that, for $\gamma=2$, the potential is constant and the
perfect-fluid term is that of stiff matter as we expect for the kinetic part
of the scalar field. Furthermore, for $\gamma=1$, we have the UDM scalar field
potential which has been found before \cite{berta,nunes}. The difference
between this solution and that of \cite{palia1} is that the free parameters
have been selected so that the stiff fluid of the kinetic part of the field is
eliminated. \ \ The transformation between the two line elements (\ref{SF.12})
and (\ref{SF.1}) is%
\begin{equation}
\omega=\frac{4}{\gamma}\ln\left(  \frac{1}{\Lambda}\exp\left(  \frac{9}%
{2}\gamma^{2}\Lambda H_{0}^{2}t^{2}\right)  -\frac{\Omega_{m0}}{2}\right)
-9\gamma\Lambda H_{0}^{2}t^{2}-\frac{4\ln\left(  2\right)  }{\gamma}.
\label{sd.12}%
\end{equation}

\subsection{Exponential function}

Assume now that $F\left(  \omega\right)  $ is an exponential function,
$\ F\left(  \omega\right)  =2F_{0}e^{F_{1}\omega}$, which gives that%
\begin{equation}
H^{2}\left(  a\right)  =\frac{1}{6}\exp\left(  -F_{0}a^{6F_{1}}\right)
\label{sd.51}%
\end{equation}
while for the scalar field we find that%
\begin{equation}
\phi\left(  \omega\right)  =\frac{2\sqrt{3F_{0}F_{1}}}{3}e^{\frac{F_{1}}%
{2}\omega}~,~V\left(  \omega\right)  =\frac{\exp\left(  -2F_{0}e^{F_{1}\omega
}\right)  }{12}\left(  1-2F_{0}F_{1}e^{F_{1}\omega}\right)  , \label{sd.52}%
\end{equation}
which gives the potential%
\begin{equation}
V\left(  \phi\right)  =\frac{1}{24}e^{-\frac{3}{2}F_{1}\phi^{2}}\left(
2-3\left(  F_{1}\right)  ^{2}\phi^{2}\right)  . \label{sd.53}%
\end{equation}

Finally, the parameter of the equation of state for the scalar field is%
\begin{equation}
w_{\phi}=-1+4F_{0}F_{1}e^{F_{1}\omega}, \label{sd.54}%
\end{equation}
and after the transformation $d\omega\rightarrow dt$ gives this in terms of
the inverse function of the exponential integral.

\subsection{Chaplygin gas}

Suppose that the scalar field satisfies the barotropic equation for the
Chaplygin gas \cite{gas}, that is,\footnote{Note that in a flat FLRW universe
the Chaplygin gas is simply a bulk viscous stress for a pressurefree fluid
with a bulk viscous coefficient proportional to $\rho^{-3/2}$. Similarly, the
generalised Chaplygin gas with $p\propto\rho^{\mu}$ is simply a bulk viscous
stress proportional to $\rho^{\mu+1/2}.$ The bulk viscous solutions that
correspond to all the Chaplygin gas models can therefore be found in ref
\cite{jdb1988}.}
\begin{equation}
p_{\phi}=\frac{A_{0}}{144}\left(  \rho_{\phi}\right)  ^{-1}. \label{sd.13}%
\end{equation}
When we substitute from (\ref{so.03}) and solve the first-order differential
equation, we find%
\begin{equation}
F\left(  \omega\right)  =-\ln\left(  \sqrt{A_{1}e^{-\omega}-A_{0}}\right)  ,
\label{sd.14}%
\end{equation}
where $A_{1}$ is a constant of integration. Therefore, we have%
\begin{equation}
\phi^{2}\left(  \omega\right)  =\frac{1}{3}\text{\textrm{arctanh}}^{2}\left(
\sqrt{1-\frac{A_{0}}{A_{1}}e^{\omega}}\right)  \label{sd.15}%
\end{equation}
and%
\begin{equation}
V\left(  \omega\right)  =\frac{1}{24}\frac{\left(  2A_{0}e^{\omega/2}%
-A_{1}e^{-\omega/2}\right)  }{\sqrt{A_{1}-A_{0}e^{\omega}}}, \label{sd.16}%
\end{equation}
which gives%
\begin{equation}
V\left(  \phi\right)  =\frac{\sqrt{A_{0}}}{24}\sinh\left(  \sqrt{3}%
\phi\right)  \left(  2-\coth\left(  \sqrt{3}\phi\right)  \right)  .
\label{sd.17}%
\end{equation}

Furthermore, for the parameter of the equation of state, we have%
\begin{equation}
w_{\phi}\left(  \omega\right)  =\frac{A_{0}e^{\omega}}{A_{1}-A_{0}e^{\omega}},
\label{sd.18}%
\end{equation}
while the transformation $d\omega\rightarrow dt$ now gives this in terms of
the inverse hypergeometric function.

\subsection{Generalized Chaplygin gas I}

The first generalization of the Chaplygin gas is by a modification of the
equation of state to \cite{bento,akam}
\begin{equation}
p_{\phi}=12^{\mu}A_{0}\left(  \rho_{\phi}\right)  ^{\left(  \mu+1\right)
},\label{sd.19}%
\end{equation}
where for $\mu=0$ we are in the limit of a perfect fluid, and for $\mu=-2$ we
have a Chaplygin gas (\ref{sd.13}).

For the function $F\left(  \omega\right)  $ we find
\begin{equation}
F\left(  \omega\right)  =\frac{1}{\mu}\ln\left(  A_{1}e^{\frac{\mu}{2}\omega
}-A_{0}\right)  .
\end{equation}
For the scalar field it follows that%
\begin{equation}
\phi\left(  \omega\right)  =\frac{2\sqrt{3}}{3\mu}\text{\textrm{arctanh}%
}\left(  \sqrt{1-\frac{A_{0}}{A_{1}}e^{-\frac{\mu}{2}\omega}}\right)
\label{sd.20}%
\end{equation}
and%
\begin{equation}
V\left(  \omega\right)  =\frac{1}{24}\left(  A_{1}e^{\frac{\mu}{2}\omega
}-2A_{0}\right)  \left(  A_{1}e^{\frac{\mu}{2}\omega}-A_{0}\right)
^{-1-\frac{1}{\mu}}. \label{sd.21}%
\end{equation}
From (\ref{sd.20}) and (\ref{sd.21}) the potential $V\left(  \phi\right)  $ is
given by the following closed-form expression%
\begin{equation}
V\left(  \phi\right)  =\frac{\left(  A_{0}\right)  ^{-\frac{1}{n}}}{24}\left(
\cosh^{2}\left(  \frac{\sqrt{3}}{2}\mu^{2}\phi\right)  -2\right)  \left(
\sinh^{2}\left(  \frac{\sqrt{3}}{2}\mu^{2}\phi\right)  \right)  ^{-1-\frac
{1}{\mu}}. \label{sd.22}%
\end{equation}

Furthermore, for the equation of state parameter
\begin{equation}
w_{\phi}\left(  \omega\right)  =\frac{A_{0}}{A_{1}e^{\frac{\mu}{2}\omega
}-A_{0}}, \label{sd.23}%
\end{equation}
while the transformation $d\omega\rightarrow dt$ is expressed in terms of the
inverse hyperbolic function.

\subsection{Generalized Chaplygin gas II}

In \cite{jdbchg} a generalized Chaplygin gas was proposed with barotropic
equation
\begin{equation}
p_{\phi}=\gamma\rho_{\phi}^{\lambda}-\rho_{\phi}, \label{sd.26}%
\end{equation}
from which we can see that for $\lambda=1$ a perfect fluid is recovered, while
for $\lambda=0$ expression (\ref{sd.26}) reduces to a special form of
(\ref{sd.04}). Again, by substitution of (\ref{so.03}) into (\ref{sd.26}) we
find that the solution of the first-order differential equation is
\begin{equation}
F\left(  \omega\right)  =-\frac{1}{1-\lambda}\ln\left(  \bar{\gamma}%
\omega+\gamma_{1}\right)  , \label{sd.27}%
\end{equation}
where $\bar{\gamma}=2^{1-2\lambda}3^{1-\lambda}\left(  \lambda-1\right)
\gamma,$ and $\gamma_{1}$ is a integration constant of integration. In what
follows we assume that $\lambda\neq0,1.$

Hence, it follows that
\begin{equation}
\phi^{2}\left(  \omega\right)  =\frac{2}{3\bar{\gamma}}\frac{\left(
\bar{\gamma}\omega+\gamma_{1}\right)  }{\left(  \lambda-1\right)  }
\label{sd.28}%
\end{equation}
and%
\begin{equation}
V\left(  \omega\right)  =\frac{\left(  \bar{\gamma}\omega+\gamma_{1}\right)
^{-1+\frac{1}{1-\lambda}}\left(  \left(  \lambda-1\right)  \left(  \bar
{\gamma}\omega+\gamma_{1}\right)  -\bar{\gamma}\right)  }{12\left(
\lambda-1\right)  }. \label{sd.30}%
\end{equation}
This gives%
\begin{equation}
V\left(  \phi\right)  =\frac{1}{36\left(  \lambda^{2}-1\right)  ^{2}}\left(
\frac{3}{2}\bar{\gamma}\left(  \lambda-1\right)  \right)  ^{\frac{1}%
{1-\lambda}}\phi^{-2+\frac{1}{1-\lambda}}\left(  3\left(  \lambda-1\right)
^{2}\phi^{2}-2\right)  . \label{sd.31}%
\end{equation}

Also, from (\ref{so.03}), it follows that $\rho_{\phi}\left(  \omega\right)
=\frac{1}{12}\left(  \bar{\gamma}\omega+\gamma_{1}\right)  ^{\frac
{1}{1-\lambda}}$ and%
\begin{equation}
w_{\phi}\left(  \omega\right)  =-1+\frac{2\bar{\gamma}}{\lambda-1}\left(
\bar{\gamma}\omega+\gamma_{1}\right)  ^{-1}. \label{sd.32}%
\end{equation}

For the scale factor $a\left(  t\right)  $ in the line element (\ref{SF.1}),
we find that $\omega=-\frac{\gamma_{1}}{\bar{\gamma}}+\left(  \left(
\frac{\bar{\gamma}}{2}\frac{2\lambda-1}{\lambda-1}\right)  ^{2}\right)
^{-2\frac{1-\lambda}{2\lambda-1}}t^{-\frac{2-2\lambda}{2\lambda-1}}%
~$for$~\lambda\neq\frac{1}{2}$, while for $\lambda=\frac{1}{2}$ it follows
that $\omega=-\frac{\gamma_{1}}{\bar{\gamma}}+\frac{1}{\bar{\gamma}}%
e^{\bar{\gamma}t}.~$Therefore, for the cosmological scale factor we have
\begin{equation}
a\left(  t\right)  \simeq\exp\left(  a_{1}t^{N}\right)  ~,~\lambda\neq\frac
{1}{2}~ \label{sd.33}%
\end{equation}
and%
\begin{equation}
a\left(  t\right)  \simeq\exp\left(  a_{1}e^{\bar{\gamma}t}\right)
~,~\lambda=\frac{1}{2}. \label{sd.34}%
\end{equation}

\subsection{Generalized Chaplygin gas III}

Consider now a third modification of the Chaplygin gas in which the pressure
and the energy density for the scalar field satisfy the nonlinear relation%
\begin{equation}
p_{\phi}=A\rho_{\phi}^{\left(  2+\lambda\right)  }+B\rho_{\phi}^{-\lambda},
\label{sd.34a}%
\end{equation}
from which, for $\lambda=-1,$ we have that $p_{\phi}=\left(  A+B\right)
\rho_{\phi}$. Equation (\ref{sd.34a}) differs from that of \cite{jacobigas} by
a term $\rho_{\phi}$. Using (\ref{so.03}), we find that
\begin{equation}
F\left(  \omega\right)  =-\frac{1}{1+\lambda}\ln\left[  F_{0}\left(
F_{1}\tanh\left(  \frac{\left(  1+\lambda\right)  }{4}F_{1}\omega\right)
-1\right)  \right]  , \label{sd.34b}%
\end{equation}
where $F_{0}=6^{-\frac{1}{1+\lambda}}12^{\lambda}A^{-1}~,~F_{1}=\sqrt{1-4AB}.$

Hence, we have that
\begin{align}
\phi\left(  \omega\right)   &  =\frac{\sqrt{6}}{6\left(  1+\lambda\right)
}\sqrt{\frac{F_{1}}{1-F_{1}}}\sqrt{\cosh^{2}\left(  \frac{\left(
1+\lambda\right)  }{4}F_{1}\omega\right)  -\sinh^{2}\left(  \frac{\left(
1+\lambda\right)  }{4}F_{1}\omega\right)  }~\times\nonumber\\
&  ~~~~~~~~~\times~\mathrm{F}_{e}\left(  \left(  \frac{F_{1}}{1+F_{1}}\left(
1-\tanh\left(  \frac{\left(  1+\lambda\right)  }{4}F_{1}\omega\right)
\right)  \right)  ,-\sqrt{\frac{1+F_{1}}{1-F_{1}}}\right)  , \label{sd.35c}%
\end{align}
where $\mathrm{F}_{e}\left(  \omega,x\right)  $ is the incomplete elliptic integral.

Furthermore, for the potential we find
\begin{equation}
V\left(  \omega\right)  =\frac{\left(  F_{0}\right)  ^{\frac{1}{1+\lambda}}%
}{48}\left(  F_{1}+2\sinh\left(  \frac{F_{1}\left(  1+\lambda\right)  }%
{2}\omega\right)  \right)  \left(  F_{1}\tanh\left(  \frac{\left(
1+\lambda\right)  }{4}F_{1}\omega\right)  -1\right)  ^{\frac{1}{1+\lambda}%
-1}\left(  \frac{F_{1}-4\cosh^{2}\left(  \frac{\left(  1+\lambda\right)  }%
{4}F_{1}\omega\right)  }{\cosh^{2}\left(  \frac{\left(  1+\lambda\right)  }%
{4}F_{1}\omega\right)  }\right)  . \label{sd.36d}%
\end{equation}
Finally, the transformation $\omega\rightarrow t$ is given now in terms of
hypergeometric functions. However, in the limit of large $\omega,$ expression
(\ref{sd.34b}) becomes constant and the solution approaches the de Sitter universe.

\subsection{Generalized Chaplygin gas IV}

We now consider another generalization of the basic Chaplygin gas, with
equation of state:
\begin{equation}
p_{\phi}=\frac{1}{6}\frac{A}{B-12\rho_{\phi}}-\rho_{\phi}, \label{se.01}%
\end{equation}
which for $A=0$, reduces to the cosmological constant, and for $B=0$, to the
Chaplygin gas II model, above, with $\lambda=-1$. On the other hand, for $B=0$
and $\rho_{\phi}\rightarrow0$, the behaviour is that of the basic Chaplygin
gas (\ref{sd.13}).

From (\ref{se.01}), we have the two solutions:
\begin{equation}
F_{\pm}\left(  \omega\right)  =-\ln\left(  B\pm\sqrt{B^{2}-2A\omega}\right)  .
\label{se.02}%
\end{equation}

Without loss of generality we work with the $F_{+}$ solution, so for the
scalar field we find%
\begin{equation}
\phi\left(  \omega\right)  =\frac{B\sqrt{\bar{\omega}\left(  \bar{\omega
}-B\right)  }\ln\left(  B+2\left(  \bar{\omega}-B+\sqrt{\bar{\omega}\left(
\bar{\omega}-B\right)  }\right)  \right)  -2\bar{\omega}\left(  \bar{\omega
}-B\right)  }{\sqrt{24A\bar{\omega}\left(  \bar{\omega}-B\right)  }},
\label{se.03}%
\end{equation}%
\begin{equation}
V\left(  \omega\right)  =\frac{\bar{\omega}\left(  \bar{\omega}-B\right)
-A}{12\left(  \bar{\omega}-B\right)  }, \label{se.04}%
\end{equation}
where $2A\omega=2B\bar{\omega}-\bar{\omega}^{2}$.

For the equation of state parameter we have
\begin{equation}
w_{\phi}=\frac{\bar{\omega}\left(  \bar{\omega}-B\right)  -2A}{\bar{\omega
}\left(  \bar{\omega}-B\right)  }, \label{se.05}%
\end{equation}
and the transformation $\omega\rightarrow t$ is the real solution of the
algebraic equation, $2\left(  3B-\bar{\omega}\right)  \sqrt{\bar{\omega}%
}=3At;$ that is:%
\begin{equation}
\ln\left(  a\left(  t\right)  \right)  =\frac{\left(  2B+\left(  9A^{2}%
t^{2}-8B^{3}+3\sqrt{A^{2}t^{2}\left(  9A^{2}t^{2}-16B^{3}\right)  }\right)
^{1/3}\right)  ^{2}}{3\left(  9A^{2}t^{2}-8B^{3}+3\sqrt{A^{2}t^{2}\left(
9A^{2}t^{2}-16B^{3}\right)  }\right)  ^{1/3}}, \label{se.06}%
\end{equation}
for $\left(  9A^{2}t^{2}-16B^{3}\right)  >0$. \ From which we can see that for
large time $a\left(  t\right)  \simeq\exp\left(  t^{2}\right)  $, which is
solution of the form (\ref{sd.33}) for the Generalized Chaplygin gas I
(\ref{sd.33}). This asymptotic behaviour leads to a strong curvature
singularity as $t\rightarrow\infty$.

\subsection{Generalized Chaplygin gas V}

Let the expression for the equation of state parameter now be
\begin{equation}
p=A\rho_{\phi}^{\lambda}+B\rho_{\phi}, \label{l.1}%
\end{equation}
where, for $B=-1,~$and $A=\gamma,$ relation (\ref{sd.26}) is recovered. \ For
(\ref{l.1}) and for $B\neq-1$, we find
\begin{equation}
F\left(  \omega\right)  =\frac{1}{\lambda-1}\ln\left(  -\frac{A}{\bar{B}%
}+\frac{\exp\left(  \frac{\left(  \lambda-1\right)  }{2}\bar{B}\omega\right)
}{\bar{B}}\right)  , \label{l.2}%
\end{equation}
where $B=\bar{B}-1$. \ \ Therefore, the scalar field is%
\begin{equation}
\left(  \phi\left(  \omega\right)  \right)  ^{2}=\frac{4}{3\bar{B}\left(
\lambda-1\right)  ^{2}}\text{\textrm{arcsinh}}^{2}\left(  \frac{\exp\left(
\frac{\lambda-1}{4}\bar{B}\omega\right)  }{\sqrt{-A}}\right)  , \label{l.3}%
\end{equation}
and the potential is%
\begin{equation}
V\left(  \omega\right)  =\frac{2A+\left(  \bar{B}-2\right)  e^{\frac{\left(
\lambda-1\right)  }{2}\bar{B}\omega}}{24\left(  A-e^{\frac{\left(
\lambda-1\right)  }{2}\bar{B}\omega}\right)  }e^{-F\left(  \omega\right)  }.
\label{l.4}%
\end{equation}

From (\ref{l.3}), we have that
\begin{equation}
\exp\left(  \frac{\lambda-1}{2}\bar{B}\omega\right)  =\sinh^{2}\left(
\frac{\sqrt{3\bar{B}}\left(  \lambda-1\right)  }{2}\phi\right)  , \label{l.5}%
\end{equation}
so the potential is
\begin{equation}
V\left(  \phi\right)  =-B^{\frac{1}{1-\lambda}}\left(  2A+\left(  \bar
{B}-2\right)  \sinh^{2}\left(  \frac{\sqrt{3\bar{B}}\left(  \lambda-1\right)
}{2}\phi\right)  \right)  \left(  \sinh^{2}\left(  \frac{\sqrt{3\bar{B}%
}\left(  \lambda-1\right)  }{2}\phi\right)  -A\right)  ^{\frac{\lambda
-2}{\lambda-1}}. \label{l.6}%
\end{equation}

For the equation of state parameter we have
\begin{equation}
w_{\phi}=-1+\frac{2}{B}\frac{\exp\left(  \frac{\lambda-1}{2}\bar{B}%
\omega\right)  }{\exp\left(  \frac{\lambda-1}{2}\bar{B}\omega\right)  -A}.
\label{l.7}%
\end{equation}

The transformation linking $\omega\rightarrow t$ is given in terms of the
inverse hyperbolic function, except when $\lambda=\frac{1}{2}$, which yields
$\omega=\frac{4}{\bar{B}}\ln\left(  \frac{1+e^{-\frac{A}{4}t}}{A}\right)  $,
and so%
\begin{equation}
a\left(  t\right)  =\left(  \frac{1+e^{-\frac{A}{4}t}}{A}\right)  ^{\frac
{2}{3\bar{B}}},~\lambda=\frac{1}{2}\text{.} \label{l.8}%
\end{equation}

\subsubsection{ Bulk viscosity}

The standard kinetic model for bulk viscosity in an isotropic an homogeneous
cosmology (see for example ref. \cite{wein}) replaces the pressure $p$ by
$p^{\prime}$ where%

\begin{equation}
p^{\prime}=p-3H\eta\label{01}%
\end{equation}
and if we have a bulk viscous coefficient $\eta$ with $\eta=\alpha\rho^{n}%
,~$with $\alpha>0$ constant, then~%

\begin{equation}
p^{\prime}=p-\sqrt{3\alpha}\rho^{n+1/2} \label{02}%
\end{equation}

In a spatially-flat FLRW universe the solutions are found by solving equation
$3H^{2}=\rho$ together with the conservation equation%

\begin{equation}
0=\dot{\rho}+3H(\rho+p^{\prime})=\dot{\rho}+\sqrt{3}\rho^{1/2}(\rho+p-\sqrt
{3}\alpha\rho^{n+1/2})\ \label{2}%
\end{equation}
Picking $p=(\gamma-1)\rho$ we can see that there are special de Sitter
solutions with $H=H_{0},~$except for the special case $n=1/2$ where the
solutions for $a(t)$ are power-law in $t.$ The exact solutions for the field
equations are given in \cite{jdb1988}. When $n>1/2$ the solution starts as de
Sitter at past infinity and approaches FRW $a=t^{2/3\gamma}$ as $t\rightarrow
\infty$. The behaviour displaying approach to de Sitter as $t\rightarrow
-\infty$ does not persist when curvature, anisotropy or another non-viscous
fluid is added to the Friedmann equation. Finally, to set up the
correspondence with equation \ref{l.1} we have to identify $B\rho$ with
$(\gamma-1)\rho$ and $A$ with $-\sqrt{3\alpha}$ and $\lambda$ with $n+1/2$.

\subsection{Lambert function I}

Suppose that $F\left(  \omega\right)  $ is given by a function of the Lambert
function, $W\left(  \omega\right)  $, specifically:%
\begin{equation}
F\left(  \omega\right)  =2\ln\left(  \frac{1}{6p}\frac{W\left(  e^{\frac
{\omega}{6p}}\right)  }{W\left(  e^{\frac{\omega}{6p}}\right)  +1}\right)
\label{sd.35}%
\end{equation}
which gives that the scale factor in the line element (\ref{SF.1}) as the
simple function
\begin{equation}
a\left(  t\right)  =\left(  t\exp\left(  t\right)  \right)  ^{p}.
\label{sdd.36}%
\end{equation}

From (\ref{so.01})-(\ref{so.03}) we find that
\begin{equation}
\phi\left(  \omega\right)  =\sqrt{2p}\ln\left[  W\left(  e^{\frac{\omega}{6p}%
}\right)  \right]  \sqrt{1+W\left(  e^{\frac{\omega}{6p}}\right)  },
\label{sd.37}%
\end{equation}%
\begin{equation}
V\left(  \omega\right)  =3p^{2}+p\frac{1-3p+6pW\left(  e^{\frac{\omega}{6p}%
}\right)  }{\left(  W\left(  e^{\frac{\omega}{6p}}\right)  \right)  ^{2}}
\label{sd.38}%
\end{equation}
and%
\begin{equation}
w_{\phi}\left(  \omega\right)  =-1+\frac{2}{3p\left(  W\left(  e^{\frac
{\omega}{6p}}\right)  +1\right)  ^{2}}. \label{sd.39}%
\end{equation}

\bigskip Expressions (\ref{sd.37}) and (\ref{sd.38}) give
\begin{equation}
V\left(  \phi\right)  =p\left(  3p\left(  e^{-\frac{\sqrt{2}}{p}\phi
}+1\right)  ^{2}-e^{-\frac{\sqrt{2}}{p}\phi}\right)  \label{sd.40}%
\end{equation}
while for the $\phi\left(  t\right)  $ we have%
\begin{equation}
\phi\left(  t\right)  =\sqrt{2p}\sqrt{\left(  1+t^{p}\right)  }\ln t.
\label{sd.41}%
\end{equation}

\subsection{Lambert function II}

We select a universe (\ref{SF.1}) in which the scale factor is given from the
following formula
\begin{equation}
a\left(  t\right)  =t^{q}\exp\left(  pt\right)  , \label{sd.42}%
\end{equation}
where in general $q\neq p$, while for $p=q$ we reduce to the previous case.

We perform the transformation $t\rightarrow\omega$ in order to write the line
element in the form of (\ref{SF.12}) and find that
\begin{equation}
F\left(  \omega\right)  =2\ln\left(  \frac{W\left(  \frac{p}{q}e^{\frac
{\omega}{6q}}\right)  }{6p\left(  W\left(  \frac{p}{q}e^{\frac{\omega}{6q}%
}\right)  +1\right)  }\right)  ; \label{sd.43}%
\end{equation}
that is,
\begin{equation}
\phi\left(  \omega\right)  =\sqrt{2q}\ln\left(  W\left(  \frac{p}{q}%
e^{\frac{\omega}{6q}}\right)  \right)  \sqrt{1+W\left(  \frac{p}{q}%
e^{\frac{\omega}{6q}}\right)  }, \label{sd.44}%
\end{equation}%
\begin{equation}
V\left(  \omega\right)  =3p^{2}+\frac{p^{2}}{q}\frac{3q-1+6qW\left(  \frac
{p}{q}e^{\frac{\omega}{6q}}\right)  }{\left[  W\left(  \frac{p}{q}%
e^{\frac{\omega}{6q}}\right)  \right]  ^{2}} \label{sd.45}%
\end{equation}
and%
\begin{equation}
w_{\phi}\left(  \omega\right)  =-1+\frac{2}{3q}\left(  1+W\left(  \frac{p}%
{q}e^{\frac{\omega}{6q}}\right)  \right)  ^{-2}. \label{sd.46}%
\end{equation}

Hence, we have%
\begin{equation}
V\left(  \phi\right)  =3p^{2}+\frac{p^{2}}{q}\frac{6qe^{\frac{\phi}{\sqrt{2q}%
}}+\left(  3q-1\right)  }{e^{\frac{2\phi}{\sqrt{2q}}}}, \label{sd.47}%
\end{equation}
where easily we observe that for $q=p$ \ expression (\ref{sd.40}) is recovered.

We should mention that scale factors (\ref{sdd.36}) and (\ref{sd.42}) can be
constructed under a rescaling transformation of scalar-field solutions of the
field equations ($H\rightarrow H+$ constant) from the power-law solution
$a\left(  t\right)  =t^{\beta}$ found in ref. \cite{new}. This can also be
special solution of a two-scalar field model \cite{twoSF}.

\subsection{Error function\ solution}

Assume now that the pressure and the energy density for the scalar field
satisfy the equation of state parameter%
\begin{equation}
p_{\phi}=-\frac{A}{6}e^{-12B\rho_{\phi}}-\rho_{\phi}, \label{sd.48}%
\end{equation}
where for $B>0$, when $\rho_{\phi}\rightarrow\infty$,~$p_{\phi}=-\rho_{\phi}$.
\ This gives
\begin{equation}
F\left(  \omega\right)  =-\ln\left(  \frac{1}{B}\ln\left(  AB\omega\right)
\right)  .
\end{equation}
which is a real function when $A\omega<0.$ Therefore, for the scalar field, we
find%
\begin{equation}
\phi\left(  \omega\right)  =-\frac{4}{3}D\left(  \sqrt{2^{-1}\ln\left(
AB\omega\right)  }\right)  ,
\end{equation}
where $D(..)$, is the Dawson integral\footnote{The Dawson integral function is
$D(x)=\frac{\sqrt{\pi}}{2}\exp[-x^{2}]\operatorname{erf}$i$(x)=\exp
[-x^{2}]\int_{0}^{x}\exp[y^{2}]dy.$}.

For the scalar field potential, it follows%
\begin{equation}
V\left(  \omega\right)  =\frac{1}{12B}\left(  \frac{1}{\omega}+\ln\left(
AB\omega\right)  \right)  ,
\end{equation}
which has a minimum at $\omega=1$,~and $B>0$. \ Finally the equation of state
parameter is
\begin{equation}
w_{\phi}\left(  \omega\right)  =-1-\frac{2}{\omega\ln\left(  AB\omega\right)
}.
\end{equation}

The expansion scale factor $a\left(  t\right)  $ is given in terms of the
inverse error function.

\section{Spatially-flat FLRW spacetime with matter source}

\label{exactS2}

We continue our analysis by assuming that a perfect fluid with constant
equation of state parameter,~$p_{m}=\left(  \gamma-1\right)  \rho_{m}$, is
added to the scalar field. Now, in order to find closed-form solutions for the
scalar field, equations (\ref{so.04})-(\ref{so.07}) have to be solved.

\subsection{Solution I}

First, we consider the special case in which the scalar field has a constant
equation of state parameter equal with that of the perfect fluid, i.e.,
$w_{\phi}=\left(  \gamma-1\right)  $. From (\ref{so.06})-(\ref{so.07}) it
follows that $2F^{\prime}-\gamma=0,~$which gives
\begin{equation}
F\left(  \omega\right)  =\frac{\gamma}{2}\omega+F_{0}, \label{sd.55}%
\end{equation}
and using (\ref{so.04}) we have that $\phi\left(  \omega\right)  $ is a linear function.

For the scalar-field potentials, we derive%
\begin{equation}
V\left(  \phi\right)  =V_{0}\exp\left(  -\frac{\sqrt{3}\gamma}{\sqrt
{\gamma\left(  1-12\rho_{0}e^{F_{0}}\right)  }}\phi\right)  , \label{sd.56}%
\end{equation}
which is just the exponential potential, as expected \cite{copdyns}. In
addition we have $V_{0}=V_{0}\left(  \gamma,\rho_{m0},F_{0}\right)  $ or,
specifically,%
\begin{equation}
V_{0}=\left(  2-\gamma\right)  \exp\left(  e^{-F_{0}}\right)  +\frac{1}%
{2}\gamma\rho_{m0}. \label{sd.57}%
\end{equation}

\subsection{Solution II}

Let the scalar field have a constant equation of state parameter $w_{\phi
}=\gamma_{\phi}-1$, but in contrast to above: $\gamma_{\phi}\neq\gamma$. This
scaling solution has been studied before in \cite{rubano}. Therefore, for this
ansatz we find that the unknown function,~$F\left(  \omega\right)  ,$ of the
line element (\ref{SF.12}) has the form%
\begin{equation}
F\left(  \omega\right)  =\frac{\gamma_{\phi}}{2}\omega+\frac{\gamma
-\gamma_{\phi}}{2}\bar{F}_{0}-\ln\left(  12\rho_{m0}\exp\left(  \frac
{\gamma-\gamma_{\phi}}{2}\left(  \bar{F}_{0}-\omega\right)  \right)
-1\right)  \label{sd.58}%
\end{equation}
where we see that the linear function (\ref{sd.55}) is recovered for
$\gamma=\gamma_{\phi}.$

For the scalar field, we find that%
\begin{equation}
\phi\left(  \omega\right)  =-\frac{2\sqrt{3\gamma_{\phi}}}{3}%
\text{\textrm{arctanh}}\left(  \sqrt{12\rho_{m0}\exp\left(  \frac
{\gamma-\gamma_{\phi}}{2}\left(  \bar{F}_{0}-\omega\right)  \right)
-1}\right)  ; \label{sd.59}%
\end{equation}
that is,%
\begin{equation}
\omega=-\bar{F}_{0}+\frac{2}{\gamma-\gamma_{\phi}}\ln\left(  \frac{1+\tanh
^{2}\left(  \frac{1}{2}\sqrt{3\gamma_{\gamma}}\left(  \gamma-\gamma_{\phi
}\right)  \phi\right)  }{12\rho_{m0}}\right)  . \label{sd.60}%
\end{equation}

Furthermore, for the potential of the scalar field we find that in terms of
$\omega$ it is expressed as
\begin{equation}
V\left(  \omega\right)  =V_{0}\left(  \gamma,\gamma_{\phi},\rho_{m0},\bar
{F}_{0}\right)  \exp\left(  -\frac{\gamma_{\phi}}{2}\omega\right)
\label{sd.61}%
\end{equation}
or in terms of $\phi$ with the use (\ref{sd.60})%
\begin{equation}
V\left(  \phi\right)  =V_{0}\left(  \gamma,\gamma_{\phi},\rho_{m0},\bar{F}%
_{0}\right)  \left(  12\rho_{m0}\right)  ^{\frac{\gamma_{\phi}}{\gamma
-\gamma\phi}}e^{\bar{F}_{0}\frac{\gamma_{\phi}}{2}}\left(  1+\tanh^{2}\left(
\frac{1}{2}\sqrt{3\gamma_{\gamma}}\left(  \gamma-\gamma_{\phi}\right)
\phi\right)  \right)  ^{\frac{\gamma_{\phi}}{\gamma-\gamma\phi}} \label{sd.62}%
\end{equation}
in which
\begin{equation}
V_{0}\left(  \gamma,\gamma_{\phi},\rho_{m0},\bar{F}_{0}\right)  =\left(
\rho_{m0}+\frac{1}{24}\left(  \gamma_{\phi}-2\right)  \exp\left(  \frac
{\gamma_{\phi}-\gamma}{2}\bar{F}_{0}\right)  \right)  . \label{sd.63}%
\end{equation}

\subsection{Solution III}

We consider that the scalar provides two fluid terms: a fluid with constant
equation of state parameter $\bar{\gamma}$, and a component which mimics the
perfect fluid $\rho_{m}$. That means that we assume the Hubble function to be
\begin{equation}
H\left(  a\right)  =H_{0}\sqrt{\Omega_{1}a^{-3\gamma}+\Omega_{2}%
a^{-3\bar{\gamma}}}, \label{sd.64}%
\end{equation}
and so
\begin{equation}
F\left(  \omega\right)  =-\ln\left(  36\Omega_{1}H_{0}^{2}e^{-\frac{\gamma
}{^{2}}\omega}+36\Omega_{2}H_{0}^{2}e^{-\frac{\bar{\gamma}}{2}\omega}\right)
. \label{sd.65}%
\end{equation}

Hence, for the scalar field it follows that%
\begin{equation}
\phi\left(  \omega\right)  =\sqrt{6}\int\sqrt{\frac{\left(  \Omega
_{1}-36\gamma\Omega_{m0}\right)  e^{-\frac{\gamma}{^{2}}\omega}+\Omega
_{2}H_{0}^{2}a^{-\frac{\bar{\gamma}}{2}\omega}}{\Omega_{1}H_{0}^{2}%
e^{-\frac{\gamma}{^{2}}\omega}+\Omega_{2}H_{0}^{2}a^{-\frac{\bar{\gamma}}%
{2}\omega}}}d\omega\label{sd.66}%
\end{equation}
and%
\begin{equation}
V\left(  \omega\right)  =\frac{3}{2}\left(  \bar{\gamma}-2\right)  \Omega
_{2}H_{0}^{2}e^{-\frac{\bar{\gamma}}{2}\omega}+\frac{12\gamma\rho
_{m0}+36\left(  2-\gamma\right)  \Omega_{1}H_{0}^{2}}{24}e^{-\frac{\gamma}%
{2}\omega}, \label{sd.67}%
\end{equation}
where $\Omega_{m0}=\frac{\rho_{m0}}{3H_{0}^{2}}$. \ In order to specify the
exact form of $V\left(  \phi\right)  ,$ the inverse function $\omega\left(
\phi\right)  $ has to be determined from the integral (\ref{sd.66}). However,
for the specific case in which $\bar{\gamma}=0$, where the extra fluid term is
that of the cosmological constant, we find that the scalar field potential has
the form%
\begin{equation}
V\left(  \phi\right)  =V_{1}+\bar{V}_{0}\left(  1+V_{3}\tanh^{2}\left(
\bar{V}_{2}\phi\right)  \right)  ^{-\frac{4}{\gamma}}, \label{sd.68}%
\end{equation}
where $V_{0-2}=V_{0-3}\left(  \Omega_{1},\Omega_{2},H_{0},\gamma,\rho
_{m0}\right)  $. This is different from the exponential term and differs from
(\ref{sd.63}).

For example, if we assume that $\gamma=1$, i.e., we are in the $\Lambda
$-cosmology, with $\Omega_{1}+\Omega_{2}=1,$ we have that
\begin{equation}
V\left(  \phi\right)  =3\left(  1-\Omega_{1}\right)  H_{0}^{2}+\frac{3}%
{2}\left(  \Omega_{1}+\rho_{m0}\right)  e^{-\frac{\omega}{2}} \label{sd.69}%
\end{equation}
and
\begin{equation}
V\left(  \phi\right)  =3\left(  1-\Omega_{1}\right)  H_{0}^{2}+\frac{3}%
{8}\left(  1+\frac{\rho_{m0}}{\Omega_{1}}\right)  \left(  e^{-\lambda
\Delta\phi}-\left(  1-\Omega_{1}\right)  e^{-2\lambda\Delta\phi}\right)  ^{2}
\label{sd.70}%
\end{equation}
where $\lambda=\frac{\sqrt{3\Omega_{1}H_{0}}}{\sqrt{\Omega_{1}H_{0}^{2}%
-12\rho_{m0}}}$. \ \ This is nothing other than a special case of the UDM
model \cite{berta}; that is, the UDM provides a dust component in the field
equations. This property for the UDM potential has been found earlier
\cite{basil} and also for a class of scalar-field potentials of the form
(\ref{sd.68}) in \cite{palprd2015}. On the other hand, the exponential
behaviour of the potential is expected according to the results of
\cite{nunes} because a scaling solution is an attractor in scalar-field models
when the potentials have asymptotically exponential terms.

\subsection{De Sitter Universe}

As a final case consider that the line element (\ref{SF.1}) is that of the de
Sitter universe, $a\left(  t\right)  =a_{0}e^{H_{0}t}$, which means that
$F\left(  \omega\right)  $ in (\ref{SF.12}) is a constant function,~$F\left(
\omega\right)  =F_{0}$. That can be seen as a special case of the previous
model that we studied in which the scalar field eliminates the perfect fluid,
i.e., $\Omega_{1}=0$.

Therefore, we find the potential to be%
\begin{equation}
V\left(  \phi\right)  =\frac{1}{12}e^{-F_{0}}\left(  1-\frac{\gamma^{2}}%
{3}\phi^{2}\right)  . \label{sd.71}%
\end{equation}

We end our analysis here and we recall that, if we set $\gamma=\frac{2}{3}$
and $\rho_{m0}=-3K$, then the solutions that have been presented in this
section hold also for the scalar field model in a nonflat FLRW universe
without a matter source.

\section{Inflationary slow-roll parameters}

\label{hsrpar}

In scalar-field cosmology, the parameters%
\begin{equation}
\varepsilon_{V}=\left(  \frac{V_{,\phi}}{2V}\right)  ^{2}~\,,~\eta_{V}%
=\frac{V_{,\phi\phi}}{2V}, \label{sd.72}%
\end{equation}
are called the potential slow-roll parameters (PSR) \cite{slp} and provide us
with an inflationary universe when $\varepsilon_{V}<<1$. The condition
$\eta_{V}<<1$ is also important for the duration of the inflation phase.

Alternatively, more accurate parameters which describe the inflationary phase
of the universe are provided by the so-called Hubble slow-roll parameters
(HSR) \cite{slpv}%
\begin{equation}
\varepsilon_{H}=-\frac{d\ln H}{d\ln a}=\left(  \frac{H_{,\phi}}{H}\right)
^{2}, \label{sd.73}%
\end{equation}
and
\begin{equation}
\eta_{H}=-\frac{d\ln H_{,\phi}}{d\ln a}=\frac{H_{,\phi\phi}}{H}. \label{sd.74}%
\end{equation}

The PSR parameters and HSR parameters are related exactly through the
relations
\begin{equation}
\varepsilon_{V}=\varepsilon_{H}\left(  \frac{3-\eta_{H}}{3-\varepsilon_{H}%
}\right)  ^{2} \label{sd.75}%
\end{equation}
and
\begin{equation}
\eta_{H}=\frac{\sqrt{\varepsilon_{H}}}{3-\varepsilon_{H}}\eta_{H,\phi}+\left(
\frac{3-\eta_{H}}{3-\varepsilon_{H}}\right)  \left(  \varepsilon_{H}+\eta
_{H}\right)  \label{sd.76}%
\end{equation}
or approximately through $\varepsilon_{V}\simeq\varepsilon_{H}$ and $\eta
_{V}\simeq\varepsilon_{H}+\eta_{H}$ when $\varepsilon_{H},\eta_{H}$ are both
very small. Therefore, when the closed-form solution of the field equations is
known, it is more accurate to work with the HSR parameters in order to study
the inflationary phase of the model rather than with the PSR parameters.

The analytical solution which was presented in Section \ref{gensf} can be used
to write the slow-roll parameters in terms of the function, $\omega$, or the
number of e-folds, $N_{e}$. Recall that the number of e-folds is given by the
formula
\begin{equation}
N_{e}=\int_{t_{i}}^{t_{f}}H\left(  t\right)  dt=\ln\frac{a_{f}}{a_{i}}%
=\frac{1}{6}\left(  \omega_{f}-\omega_{i}\right)  , \label{sd.77}%
\end{equation}
where $a_{f}=a\left(  t_{f}\right)  $ is the moment at which inflation ends,
$\varepsilon_{H}\left(  t_{f}\right)  =1,$ while $a_{i}=a\left(  t_{i}\right)
$ is the moment at which inflation starts. It is assumed that $N_{e}$ lies in
the interval $N_{e}\in\left[  50,60\right]  $. \ 

Therefore, the HSR parameters are%
\begin{equation}
\varepsilon_{H}=3F^{\prime}~,~\eta_{H}=3\frac{\left(  F^{\prime}\right)
^{2}-F^{\prime\prime}}{F^{\prime}}, \label{sd.78}%
\end{equation}
where either from (\ref{sd.75}) and (\ref{sd.76}) or directly from with the
use of (\ref{so.01}) and (\ref{so.02}) the PSR parameters can be derived in
terms of $\omega$. Here we comment that the PSR parameters depend always upon
a higher derivative of $F$, in contrast to the HSR parameters, $\varepsilon
_{V}=\varepsilon_{V}\left(  F^{\prime},F^{\prime\prime}\right)  $ and
$\eta_{V}=\eta_{V}\left(  F^{\prime},F^{\prime\prime},F^{\prime\prime\prime
}\right)  .$

In a similar way, the HSR expansion parameters \cite{slp} can be expressed in
terms of the function $F\left(  \omega\right)  $ and its derivative. For
example, the third-order HSR parameter is
\begin{equation}
\xi_{H}\equiv\frac{H_{\phi}H_{\phi\phi\phi}}{H^{2}}=-\frac{9\sqrt{6}}{4\left(
F^{\prime}\right)  ^{\frac{5}{2}}}\left[  \left(  F^{\prime}\right)
^{4}-\left(  3F^{\prime2}+2F^{\prime\prime}\right)  F^{\prime\prime
}+2F^{\prime}F^{\prime\prime\prime}\right]  . \label{sd.79}%
\end{equation}

Note that the spectral indices for the density perturbations, and for the
gravitational waves in the first approximation, are given in terms of the HSR
parameters by
\begin{equation}
n_{s}=1-4\varepsilon_{H}+2\eta_{H}~\ \ ~,~n_{g}=-2\varepsilon_{H},
\label{sd.80}%
\end{equation}
while the tensor to scalar ratio is $r=10\varepsilon_{H}$. Finally, the range
of the scalar spectral index is given by
\begin{equation}
n_{s}^{\prime}=2\varepsilon_{H}\eta_{H}-2\xi_{H}. \label{sd.81}%
\end{equation}

From the Planck 2015 collaboration \cite{planck2015}, we have that the above
parameters are $n_{s}=0.968\pm0.006~\mbox{\rm and}~\eta_{s}^{\prime}%
=-0.003\pm0.007,$ while the tensor to scalar ratio has a value smaller than
$0.11$, i.e., $r<0.11$. \ From these values some intervals for the HSR
parameters can be determined. For instance from $r$ it follows that
$\varepsilon_{H}\lesssim0.01$.

In what follows, we determine the HSR parameters for some of the solutions of
the Section \ref{exactS1} and compare them with the Planck data constraints.
\ Specifically, we study the following models: generalized Chaplygin gases
I-V; the Lambert function II model, and the error function solution.

\subsection{Generalized Chaplygin gas I}

For the generalized Chaplygin gas model, (\ref{sd.19}), the HSR parameters are
given by%

\begin{equation}
\varepsilon_{H}=\frac{3}{2}\frac{1}{1-\frac{A_{0}}{A_{1}}e^{-\frac{\mu}%
{2}\omega}}~,~\eta_{H}=\frac{1}{2}\left(  2\varepsilon_{H}\left(
1+\mu\right)  -3\mu\right)  , \label{sd.87}%
\end{equation}
and
\begin{equation}
\xi_{H}=-\frac{3\sqrt{2}}{8}\sqrt{\varepsilon_{H}}\left[  \left(
1+\mu\right)  \left(  1+2\mu\right)  \sqrt{\varepsilon_{H}}-3\mu\left(
3+2\mu\right)  \right]  , \label{sd.88}%
\end{equation}
where, for $\mu=-2$, the parameters reduce to those of the basic Chaplygin
gas, (\ref{sd.13}). Moreover, inflation ends when $\omega_{f}=-\frac{2}{\mu
}\ln\left(  -\frac{1}{2}\frac{A_{1}}{A_{0}}\right)  ,~$from which we find that%
\begin{equation}
\varepsilon_{H}\left(  \omega_{i}\right)  =\frac{3}{2+e^{3N\mu}}.
\label{sd.88a}%
\end{equation}
\ \ \ ~As above, the spectral indices can be expressed in terms of
$\varepsilon_{H}$ by
\begin{equation}
n_{s}=1+2\varepsilon_{H}\left(  \mu-1\right)  -3\mu
\end{equation}
and%
\begin{equation}
n_{s}^{\prime}=2\left(  1+\mu\right)  \varepsilon_{H}^{2}-\frac{3\sqrt{2}}%
{4}\left(  2\sqrt{2}\mu-1-3\mu-2\mu^{2}\right)  \varepsilon_{H}-\frac
{9\sqrt{2}}{4}\left(  3+2\mu\right)  \sqrt{\varepsilon_{H}}.
\end{equation}

\begin{figure}[t]
\includegraphics[scale=0.55]{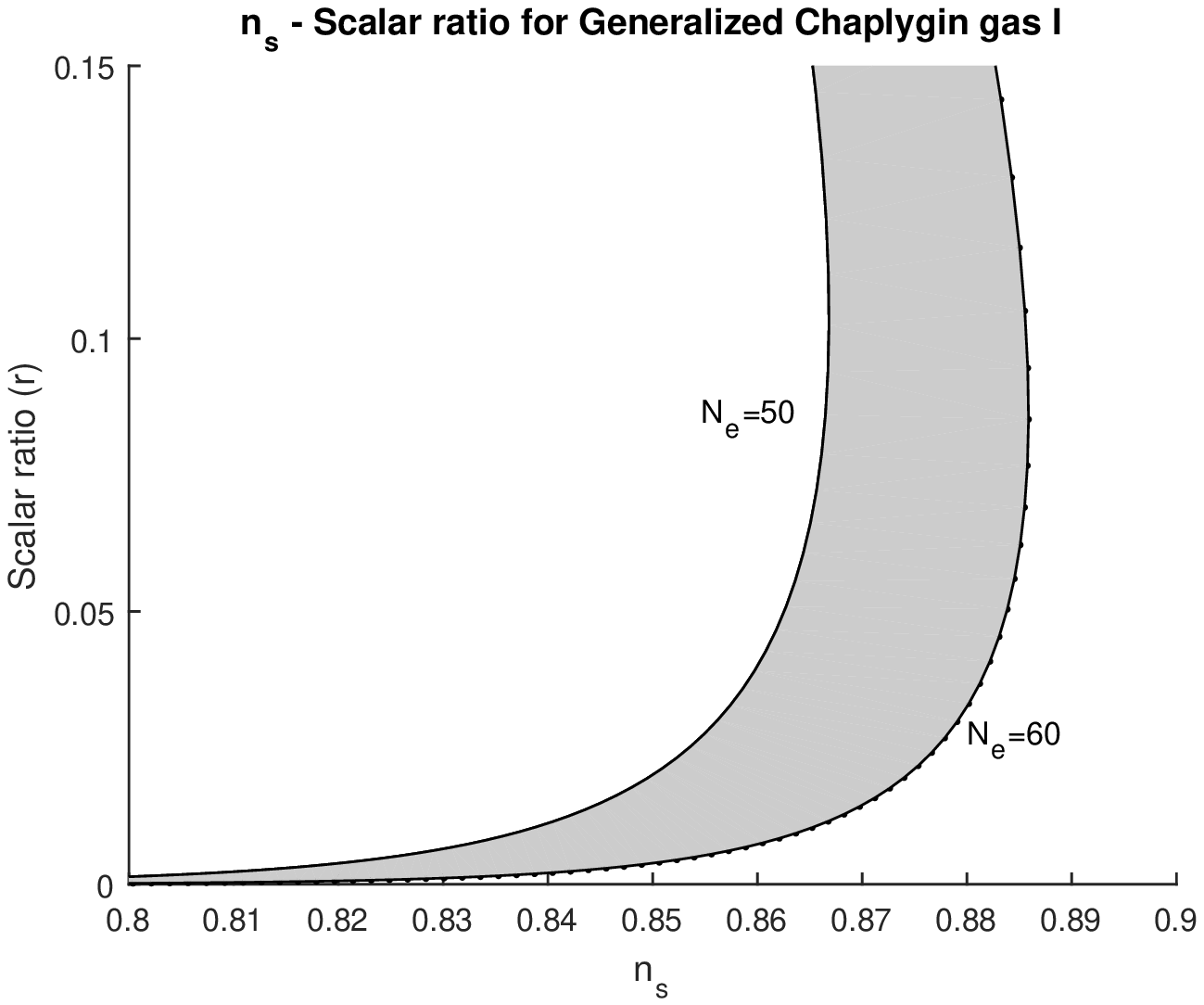}
\centering
\includegraphics[scale=0.55]{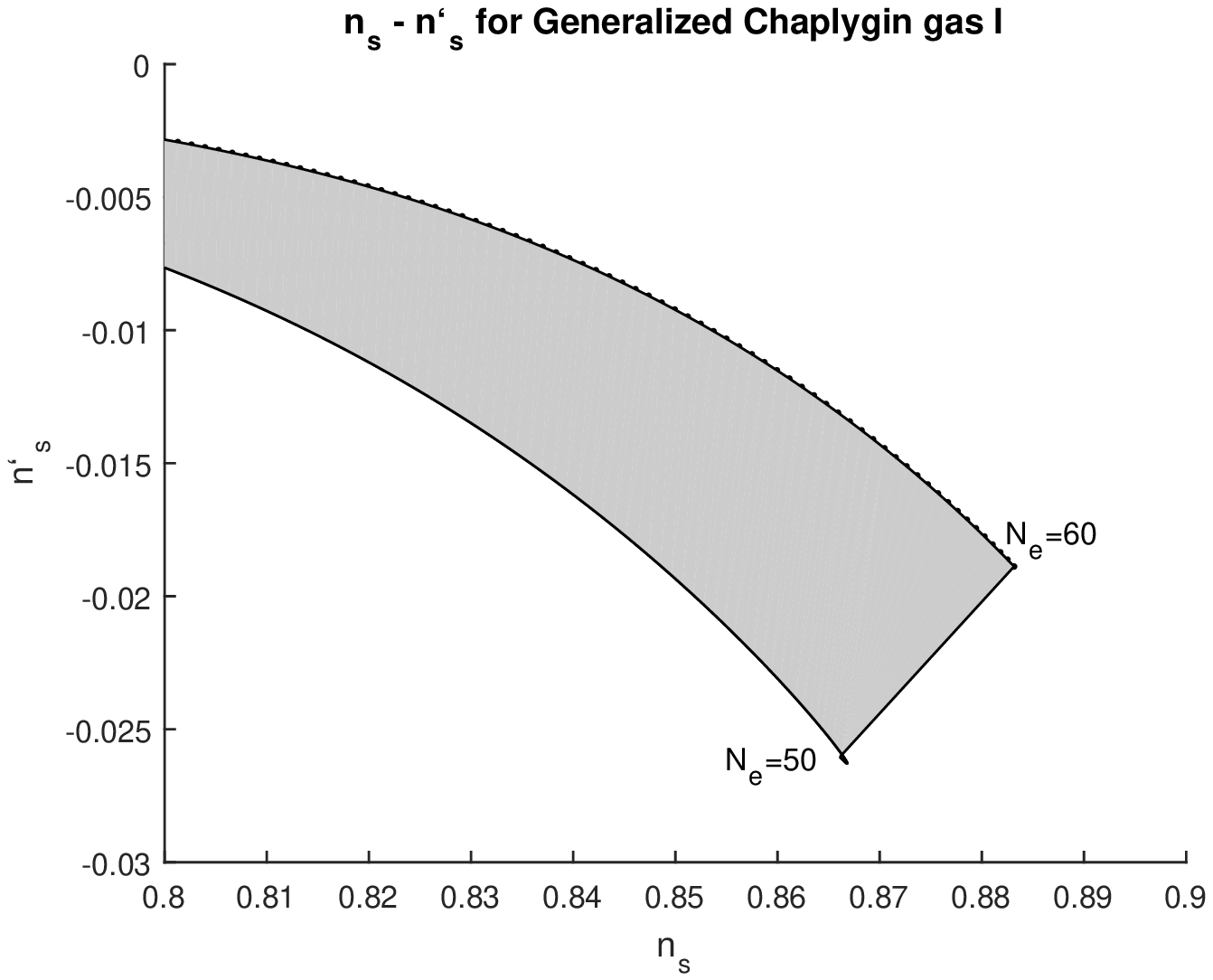}
\caption{Diagrams $r=r\left(  n_{s}\right)  $ (left figure) and $n_{s}%
^{\prime}=n_{s}^{\prime}\left(  n_{s}\right)  ~$(right figure) for the
generalized Chaplygin gas I model for number of e-folds in the range $N_{e}%
\in\left[  50,60\right]  $. }%
\label{fig1}%
\end{figure}

From (\ref{sd.88a}), we observe that in order for $\varepsilon_{H}<1$, $\mu$
should be positive. In Figure \ref{fig1} the $n_{s}-r$ and $n_{s}%
-n_{s}^{\prime}$ diagrams are presented. They reveal that for $N_{e}=60$~we
have $\max n_{s}\simeq0.887<0.968$ while at the same time \thinspace
$r\simeq0.08$ and $n_{s}^{\prime}=-0.02$, which corresponds to $\mu
\simeq0.033$; that is, a small deviation from a perfect fluid. Also, we
mention that for smaller values of $N_{e}$ the maximum of $n_{s}$ is smaller,
$n_{s}^{\prime}$ is smaller, although the scalar ratio has a similar value.
For $N_{e}=55,$ we have $\max n_{s}=0.877\,,$ $r\simeq0.09$ and $n_{s}%
^{\prime}=-0.022$ for $\mu\simeq0.032$. 

\subsection{Generalized Chaplygin gas II}

\begin{figure}[t]
\centering
\includegraphics[scale=0.55]{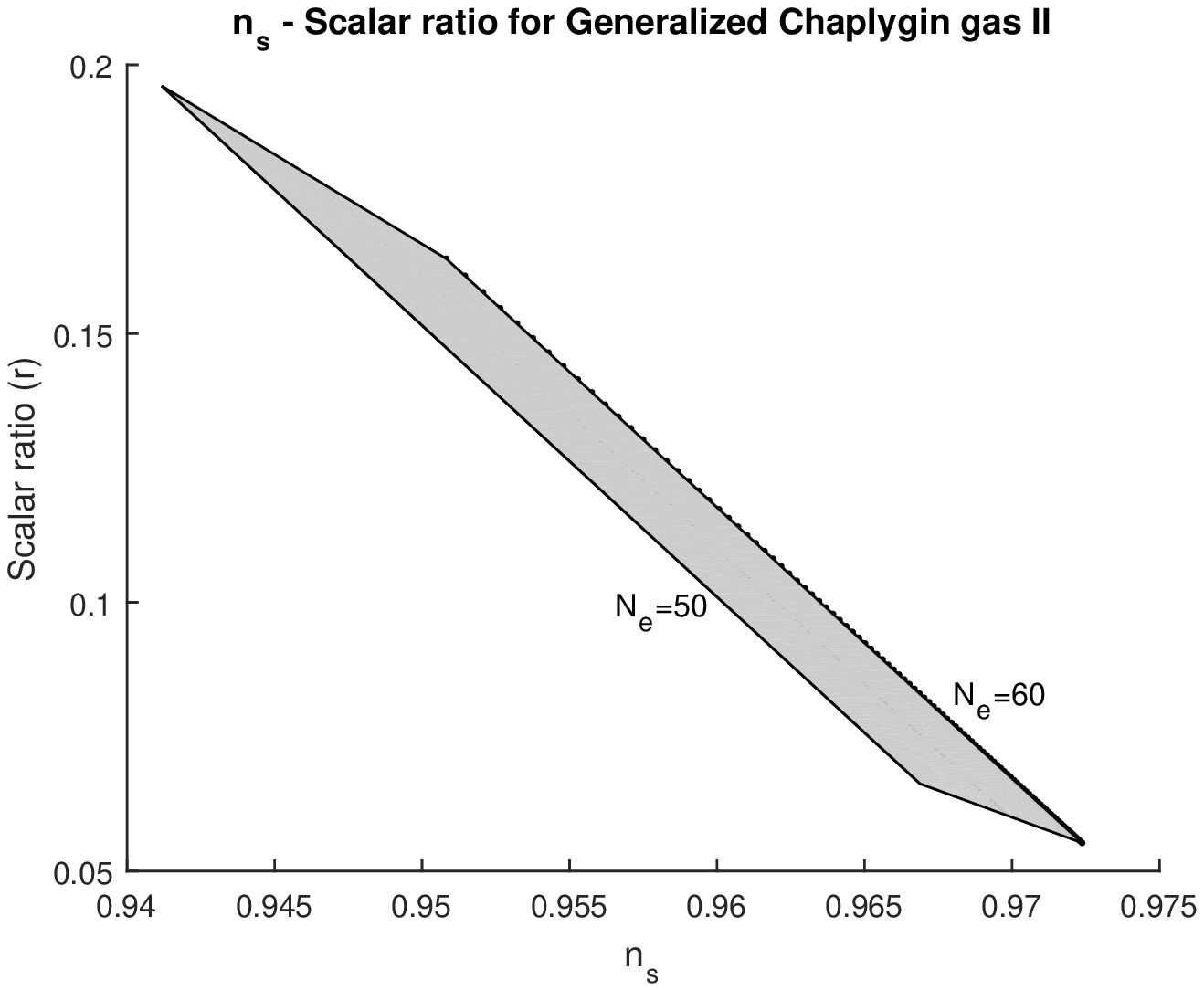}
\label{fig2a} \centering
\includegraphics[scale=0.55]{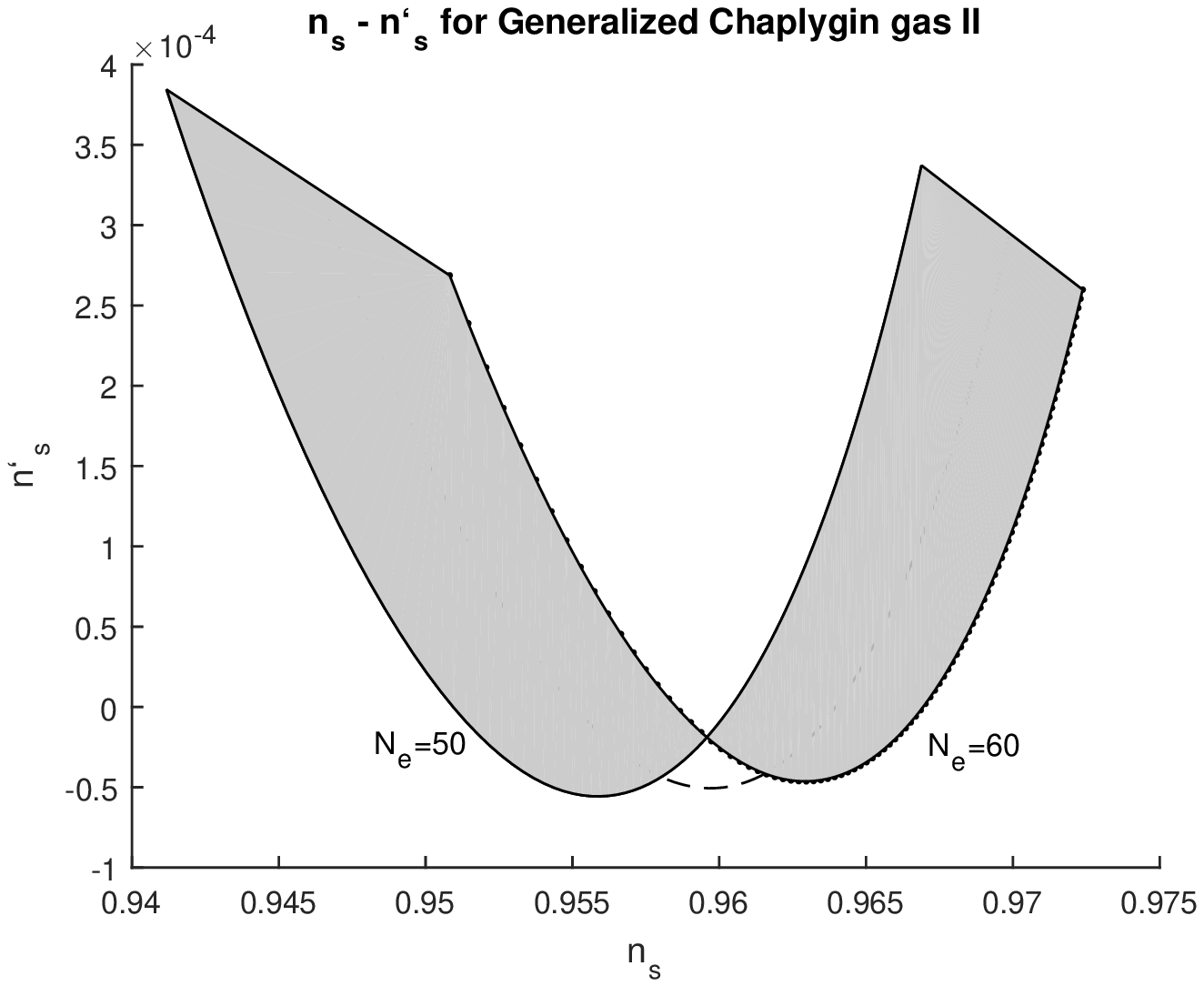}
\label{fig2b} \caption{The left figure is the $n_{s}-r$ \ diagram for the
generalized Chaplygin gas II model, while the right figure is $\ $the
$n_{s}-n_{s}^{\prime}$ diagram for the same model. The plots are for $N_{e}%
\in\left[  50,60\right]  $ and $\lambda\in\left[  -0.5,0.5\right]  $. $~$}%
\label{fig2}%
\end{figure}

Consider now the generalized Chaplygin gas II model, (\ref{sd.26}). For this
model the HSR parameters are calculated to be%

\begin{equation}
\varepsilon_{H}=\frac{3\bar{\gamma}}{\lambda-1}\left(  \bar{\gamma}%
\omega+\gamma_{1}\right)  ^{-1}~~, \label{sd.89}%
\end{equation}
and
\begin{equation}
\eta_{H}=\lambda\varepsilon_{H}~,~\xi_{H}=-\frac{\sqrt{2}}{4}\lambda\left(
2\lambda-1\right)  \left(  \varepsilon_{H}\right)  ^{\frac{3}{2}},
\label{sd.90}%
\end{equation}
where for $\lambda=\frac{1}{2}$ the parameter $\xi_{H}$ becomes zero.
\ Furthermore, we find that
\begin{equation}
n_{s}=1-4\varepsilon_{H}+2\lambda\varepsilon_{H}~,~~n_{s}^{\prime}%
=2\lambda\left(  \varepsilon_{H}\right)  ^{2}+\frac{\sqrt{2}}{2}\lambda\left(
2\lambda-1\right)  \left(  \varepsilon_{H}\right)  ^{\frac{3}{2}}.
\label{sd.90a}%
\end{equation}
and inflation ends at the point $\omega_{f}=\frac{3}{\lambda-1}-\frac
{\gamma_{1}}{\bar{\gamma}}$. Therefore, we have that
\begin{equation}
\varepsilon_{H}\left(  \omega_{i}\right)  =\left(  1+2N\left(  1-\lambda
\right)  \right)  ^{-1}, \label{sd.90b}%
\end{equation}

where $\varepsilon_{H}<1$ and $\varepsilon_{H}\geq0$ for $\lambda<1,$ while
for $\lim_{\lambda\rightarrow-\infty}\varepsilon_{H}\left(  \omega_{i}\right)
=0$. \ \ In Figure \ref{fig2} we give the $n_{s}-r$ and $n_{s}-n_{s}^{\prime}$
diagrams for various values of the parameter $\lambda$ in the range
$\lambda\in\lbrack-0.5,0.5]$. From these diagrams, we observe that for
$N_{e}=55$ (dash-dash lines), for $n_{s}=0.968$, we have~$r\simeq0.073$ and
$n_{s}^{\prime}=2\times10^{-4}$. These are values inside the range of values
consistent with the Planck 2015 collaboration, in contrast to the situation
for the Generalized Chaplygin gas I. 

\subsection{Generalized Chaplygin gas III}

For the equation of state parameter (\ref{sd.34a}), that is for the solution
(\ref{sd.34b}), we find that
\begin{equation}
\varepsilon_{H}=\frac{3\left(  F_{1}\right)  ^{2}}{4}\left(  1-\tanh\left(
\frac{\left(  1+\lambda\right)  }{4}F_{1}\omega\right)  \right)  ^{-1}%
\cosh^{-2}\left(  \frac{\left(  1+\lambda\right)  }{4}F_{1}\omega\right)
\label{sd.001}%
\end{equation}
and%
\begin{equation}
\eta_{H}=\frac{1}{2}\left(  2\varepsilon_{H}+\left(  1+\lambda\right)
\sqrt{\left(  \varepsilon_{H}\right)  ^{2}-3\varepsilon_{H}+9\left(
F_{1}\right)  ^{2}}\right)  , \label{sd.002}%
\end{equation}%
\begin{equation}
\xi_{H}=\frac{-3\sqrt{\varepsilon_{H}}}{4\sqrt{2}}\left(  \varepsilon
_{H}\left(  3+4\lambda+2\lambda^{2}\right)  -3\left(  1+\lambda\right)
^{2}+3\left(  1+\lambda\right)  \sqrt{4\left(  \varepsilon_{H}\right)
^{2}-12\varepsilon_{H}+9\left(  F_{1}\right)  ^{2}}\right)  . \label{sd.003}%
\end{equation}
Therefore, we have%
\begin{equation}
n_{s}=1-2\varepsilon_{H}+\left(  1+\lambda\right)  \sqrt{\left(
\varepsilon_{H}\right)  ^{2}-3\varepsilon_{H}+9\left(  F_{1}\right)  ^{2}}.
\label{sd.004}%
\end{equation}

From (\ref{sd.002}), we see that $\eta_{H}\rightarrow0$, when $\varepsilon
_{H}\rightarrow0$, if and only if $3\left(  1+\lambda\right)  \left\vert
F_{1}\right\vert \rightarrow0$, while at the same time $\xi_{H}\rightarrow0$,
that is $n_{s}\rightarrow1$ and $n_{s}^{\prime}\rightarrow0$. \ Furthermore,
from (\ref{sd.001}), we find that inflation ends at%
\begin{equation}
e^{\omega_{f}}=2^{-\frac{2}{\left(  1+\lambda\right)  F_{1}}}\left(
\frac{3\left(  F_{1}\right)  ^{2}-8\sqrt{9\left(  F_{1}\right)  ^{2}-8}%
-2}{F_{1}-1}\right)  ^{\frac{2}{\left(  1+\lambda\right)  F_{1}}},
\label{sd.005}%
\end{equation}
which requires $\left(  F_{1}\right)  ^{2}\geq\frac{8}{9}$; that is, from the
above, $\lambda$ should be very close to $-1$.

\begin{figure}[t]
\centering
\includegraphics[scale=0.55]{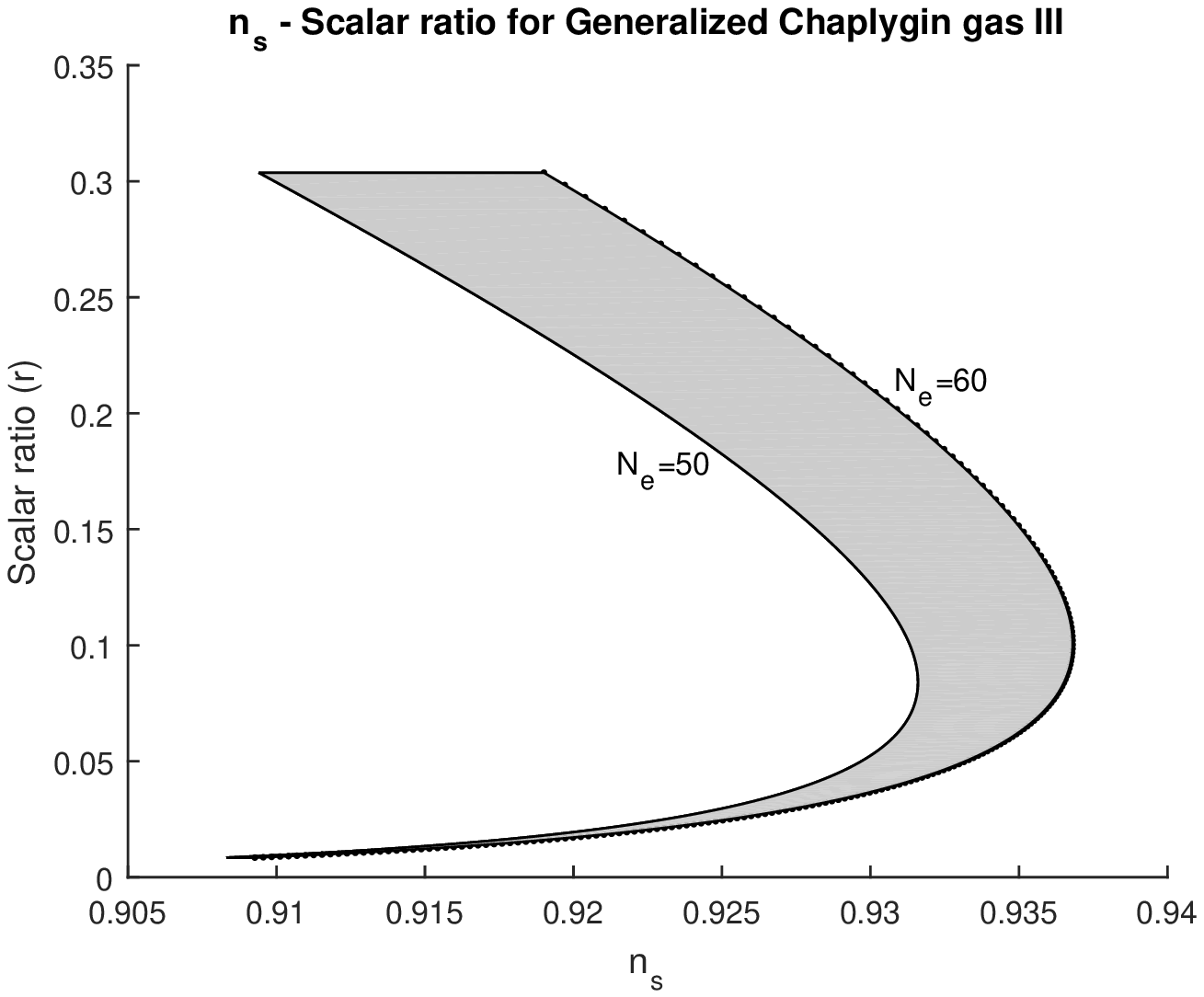}
\label{fig4a} \centering
\includegraphics[scale=0.55]{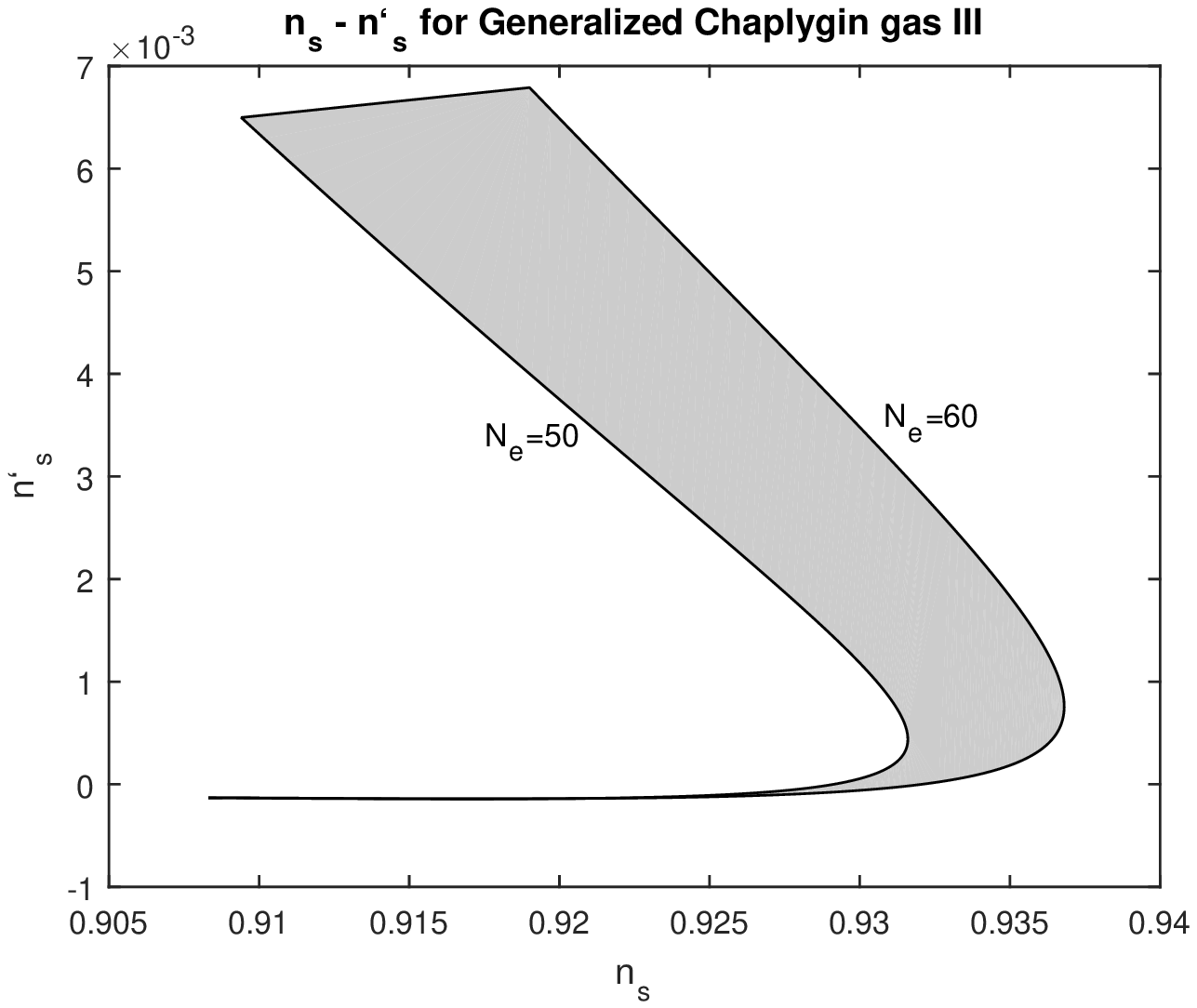}
\label{fig4b} \caption{Left figure is the $n_{s}-r$ \ diagram for the
generalized Chaplygin gas III model, while the right figure is $\ $the
$n_{s}-n_{s}^{\prime}$ diagram for the same model. The plots are for $N_{e}%
\in\left[  50,60\right]  $ and $\lambda\in\left[  -1.03,-1.01\right]  $,
$\ $with $F_{1}=-1.0001$. $~$}%
\label{fig4}%
\end{figure}

In Fig \ref{fig4}, we give the evolution of the $n_{s}-r$ and $n_{s}%
-n_{s}^{\prime}$ diagrams. We observe that $n_{s}$ reaches the observed value
$0.986$ when the number of e-folds $N_{e}$ exceeds $60$.

\subsection{Generalized Chaplygin gas IV}

\begin{figure}[t]
\centering
\includegraphics[scale=0.55]{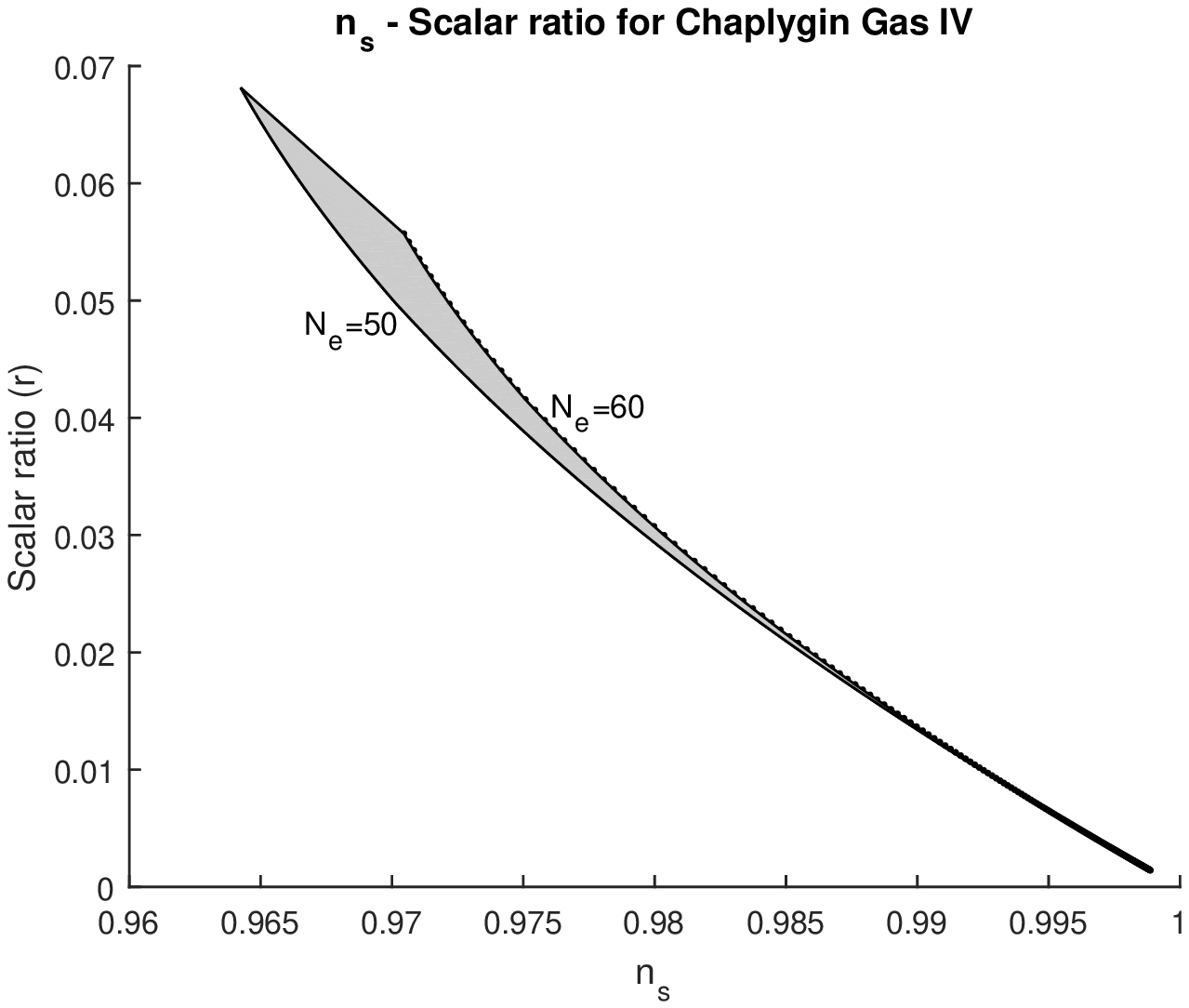}
\centering
\includegraphics[scale=0.55]{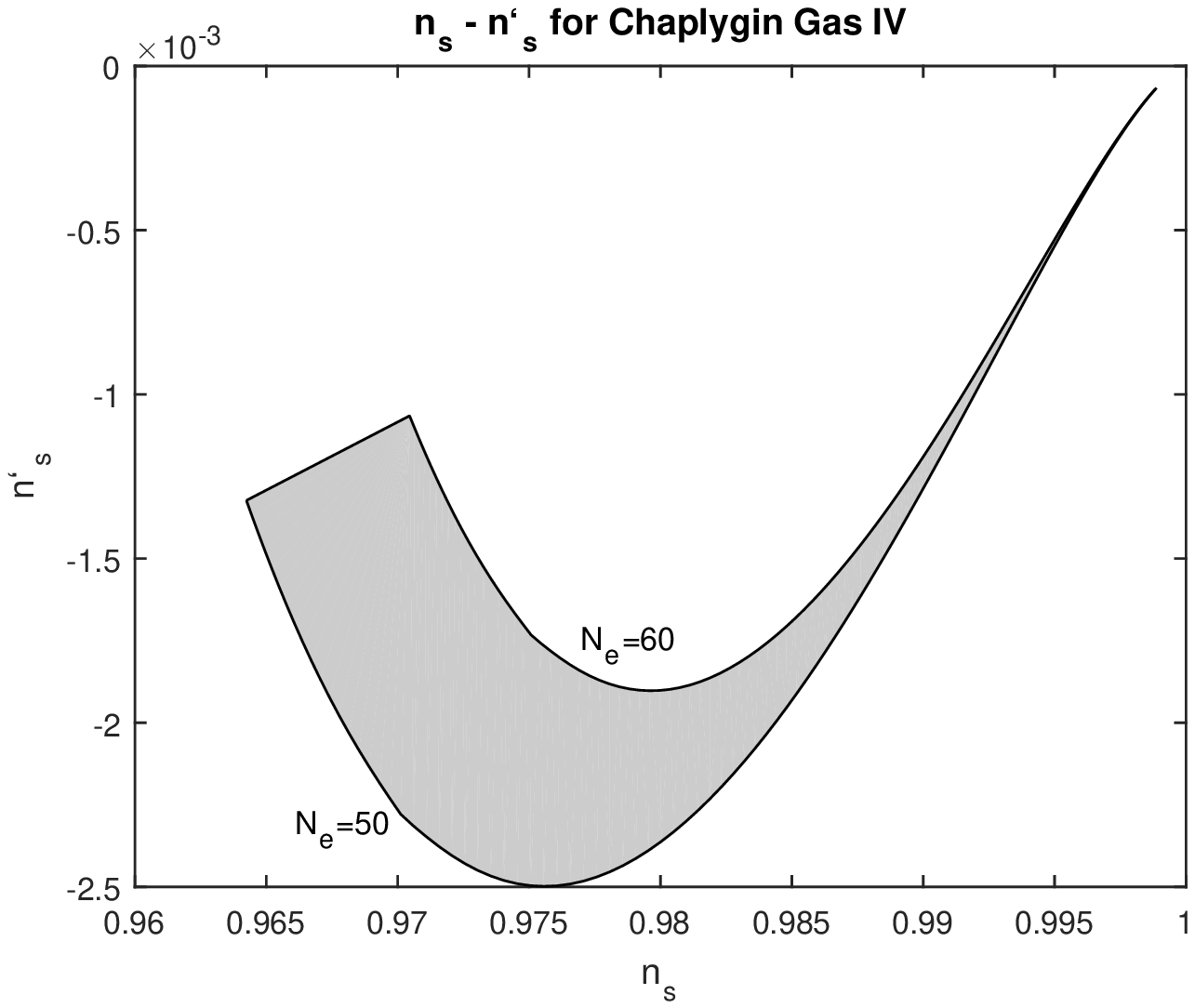}
\caption{Diagrams $r=r\left(  n_{s}\right)  $ (left figure) and $n_{s}%
^{\prime}=n_{s}^{\prime}\left(  n_{s}\right)  ~$(right figure) for the
generalized Chaplygin gas IV model for number of e-folds in the range
$N_{e}\in\left[  50,60\right]  $, for $A=1$ and free parameter $B$ in the
range $B\in\left[  -10,100\right]  $.}%
\label{fig6}%
\end{figure}

The HSR parameters for the generalized Chaplygin gas IV model (\ref{se.01})
are found to be%
\begin{equation}
\varepsilon_{H}=\frac{3A}{B^{2}-2A\omega+B\sqrt{B^{2}-2A\omega}}~,~\eta
_{H}=\frac{3A}{2A\omega-B^{2}},
\end{equation}%
\begin{equation}
\xi_{H}=-\sqrt{\frac{3}{2}}\frac{9A^{\frac{3}{2}}\left(  4B+3\sqrt
{B^{2}-2A\omega}\right)  }{2\left(  B^{2}-2A\omega\right)  ^{\frac{3}{2}%
}\left(  B\left(  B+\sqrt{B^{2}-2A\omega}\right)  -2A\omega\right)  ^{\frac
{1}{2}}},
\end{equation}
from which we find $\omega_{f}^{\pm}=\frac{B^{2}-6A\pm\sqrt{B^{2}+12AB^{2}}%
}{4A}.$ In Fig. \ref{fig6}, the $n_{s}-r$ and $n_{s}-n_{s}^{\prime}$ diagrams
are given for the $\omega_{f}^{-}$, and for various values of the parameter
$B$; we have assumed that $A=1$. From the diagrams it is easy to see that this
model can fit the Planck 2015 data quite well. Specifically, we find that for
$n_{s}\simeq0.968$ and for $N_{e}=55$, $r\simeq0.054$ and $n_{s}^{\prime
}\simeq-1.5~10^{-3}$.

\subsection{Generalized Chaplygin gas V}

\begin{figure}[t]
\centering\includegraphics[scale=0.55]{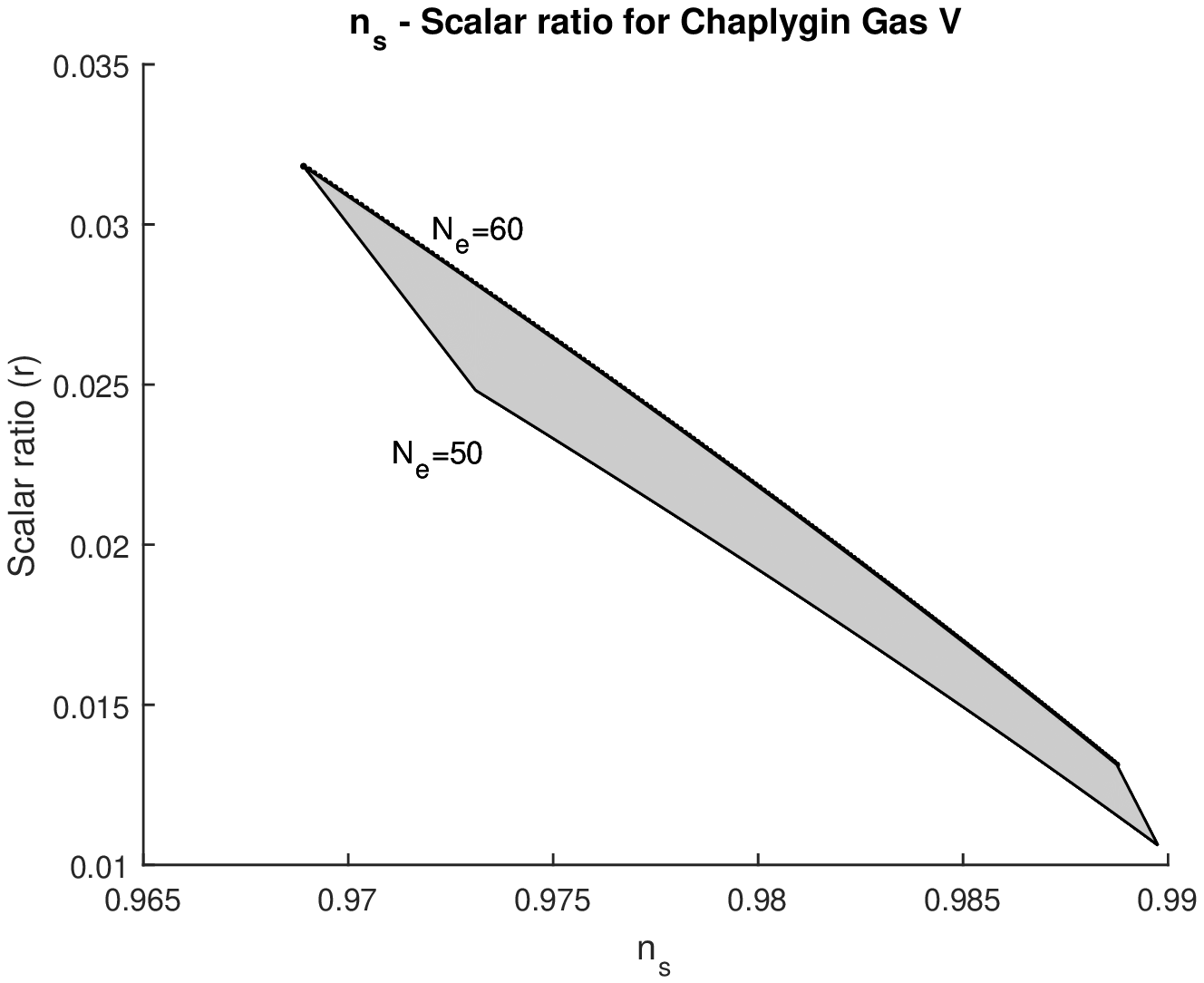}
\label{fig3aa} \centering\includegraphics[scale=0.55]{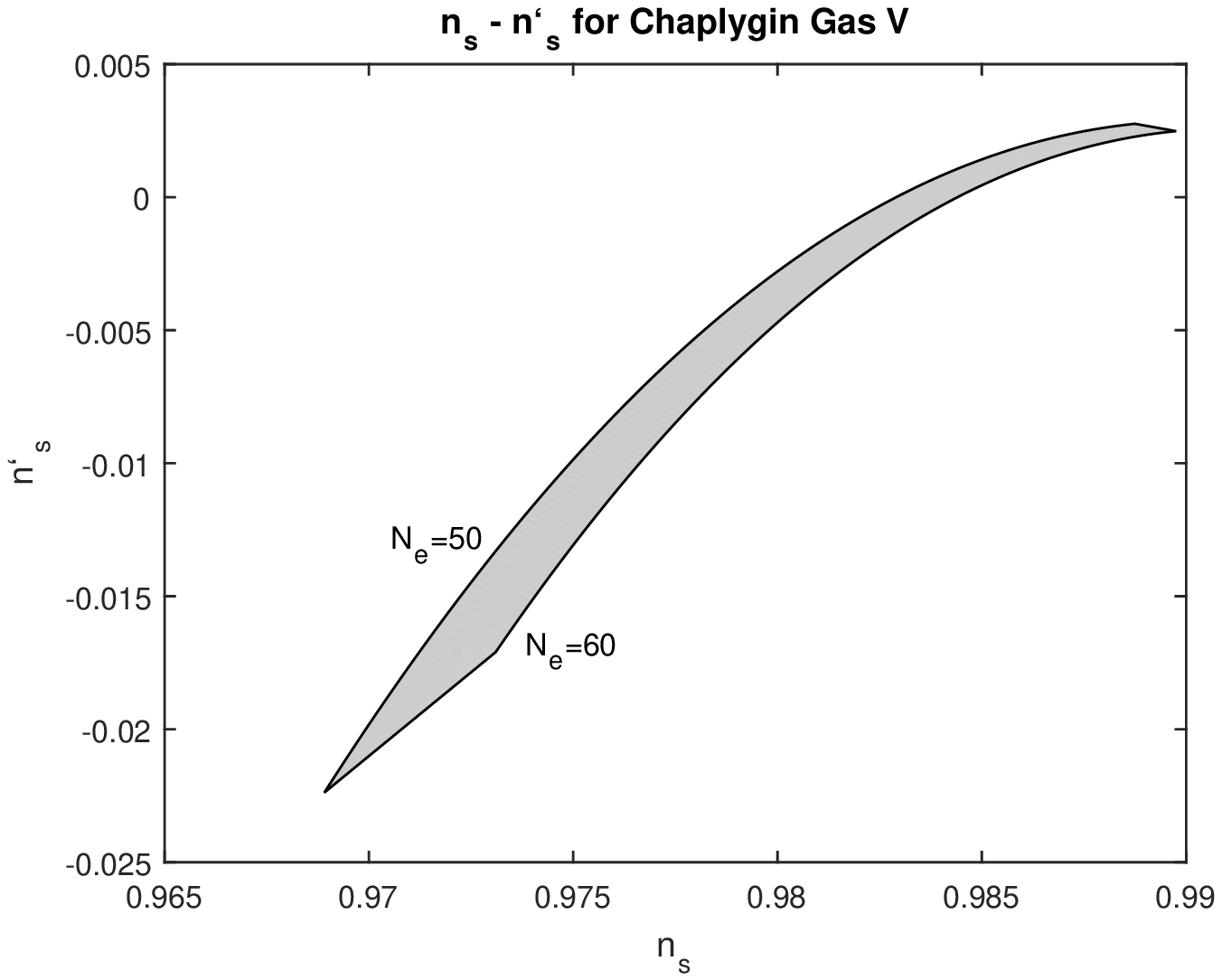}
\label{fig1ba} \caption{The left figure is the $n_{s}-r$ \ diagram for the
model generalized Chaplygin gas V. The right figure is $\ $the $n_{s}%
-n_{s}^{\prime}$ diagram for the same model. The plots are for $N_{e}%
\in\left[  50,60\right]  ,~\bar{B}=-0.002$ and $\lambda\in\lbrack-1,0)$. $~$}%
\label{fig7}%
\end{figure}

For function (\ref{l.2}), the HSR parameters are now calculated to be%
\begin{equation}
\varepsilon_{H}=\frac{3\bar{B}}{2}\left(  1-Ae^{-\frac{\lambda-1}{2}\bar
{B}\omega}\right)  ,~\eta_{H}=-\frac{3}{2}\bar{B}\left(  \lambda-1\right)
+\lambda\varepsilon_{H}, \label{l.01}%
\end{equation}
and%
\begin{equation}
\xi_{H}=-\frac{9\sqrt{2\varepsilon_{H}}}{8}\left(  \left(  \lambda-1\right)
^{2}\left(  6\bar{B}-4\varepsilon_{H}\right)  +\left(  \lambda-1\right)
\left(  9\bar{B}-6\varepsilon_{H}\right)  -2\varepsilon_{H}\right)  ,
\label{l.02}%
\end{equation}
from which we find that inflation ends when
\begin{equation}
\omega_{f}=\frac{2\left(  \ln2A-\ln\left(  2-3\bar{B}\right)  \right)  }%
{\bar{B}\left(  \lambda-1\right)  }. \label{l.03}%
\end{equation}
From (\ref{l.01}) we observe that when $\varepsilon_{H}\rightarrow0$%
,~$\eta_{H}\simeq\bar{B}\left(  \lambda-1\right)  $, hence $\eta
_{H}\rightarrow0$ when $\bar{B}\rightarrow0$, or $\lambda\rightarrow1$. Recall
that $\bar{B}=0$ means that we are in the case of the generalized Chaplygin
gas II model. \ On the other hand, by replacing $\omega_{i}=\omega_{f}-6N_{e}$
in (\ref{l.01}), (\ref{l.02}) using (\ref{l.03}), it follows that the HSR
parameters are independent on the constant $A$, and are functions of $\bar
{B},~\lambda$ and the number of e-folds $N_{e}$. \ We choose $\bar{B}=-0.002$
and for the ranges $N_{e}\in\left[  50,60\right]  $ and $\lambda\in
\lbrack-1,0)$, we present the $n_{s}-r$, and $n_{s}-n_{s}^{\prime}$ diagrams
in Fig \ref{fig7}. We can see this differs from that of the Chaplygin gas II model.

\subsection{Lambert function II}

For the scale factor (\ref{sd.42}) we find that the HSR parameters are%

\begin{equation}
\varepsilon_{H}=\frac{1}{q}\left(  1+W\left(  \frac{p}{q}e^{\frac{\omega}{6q}%
}\right)  \right)  ^{-2}~~ \label{sd.91a}%
\end{equation}
and
\begin{equation}
\eta_{H}=\sqrt{\frac{\varepsilon_{H}}{q}}~,~\xi_{H}=-\frac{3\sqrt{2}}{4q^{2}%
}\sqrt{\varepsilon_{H}}, \label{sd.92}%
\end{equation}
while the spectral indices become
\begin{equation}
n_{s}=1-4\varepsilon_{H}+2\sqrt{\frac{\varepsilon_{H}}{q}}~,~n_{s}^{\prime
}=\frac{2}{\sqrt{q}}\left(  \varepsilon_{H}\right)  ^{\frac{3}{2}}%
+\frac{3\sqrt{2}}{2q^{2}}\sqrt{\varepsilon_{H}}. \label{sd.92a}%
\end{equation}

From (\ref{sd.91a}), we find that inflation ends at $\omega_{f}=6\ln\left(
\frac{\sqrt{q}\left(  1-\sqrt{q}\right)  }{p}\right)  +\frac{1-\sqrt{q}}%
{\sqrt{q}}.$ It is important to mention here that in order for (\ref{sd.91a})
to be positive we need $q>0$, while $\omega_{f}$ is real when $\frac
{1-\sqrt{q}}{p}>0$. Furthermore, we find%
\begin{equation}
\varepsilon_{H}=\frac{1}{q}\left(  1+W\left(  \frac{\left(  1-\sqrt{q}\right)
}{\sqrt{q}}e^{\frac{1-\sqrt{q}}{\sqrt{q}}}\exp\left(  -\frac{N}{q}\right)
\right)  \right)  ^{-2}. \label{sd.93a}%
\end{equation}

\begin{figure}[t]
\centering
\includegraphics[scale=0.55]{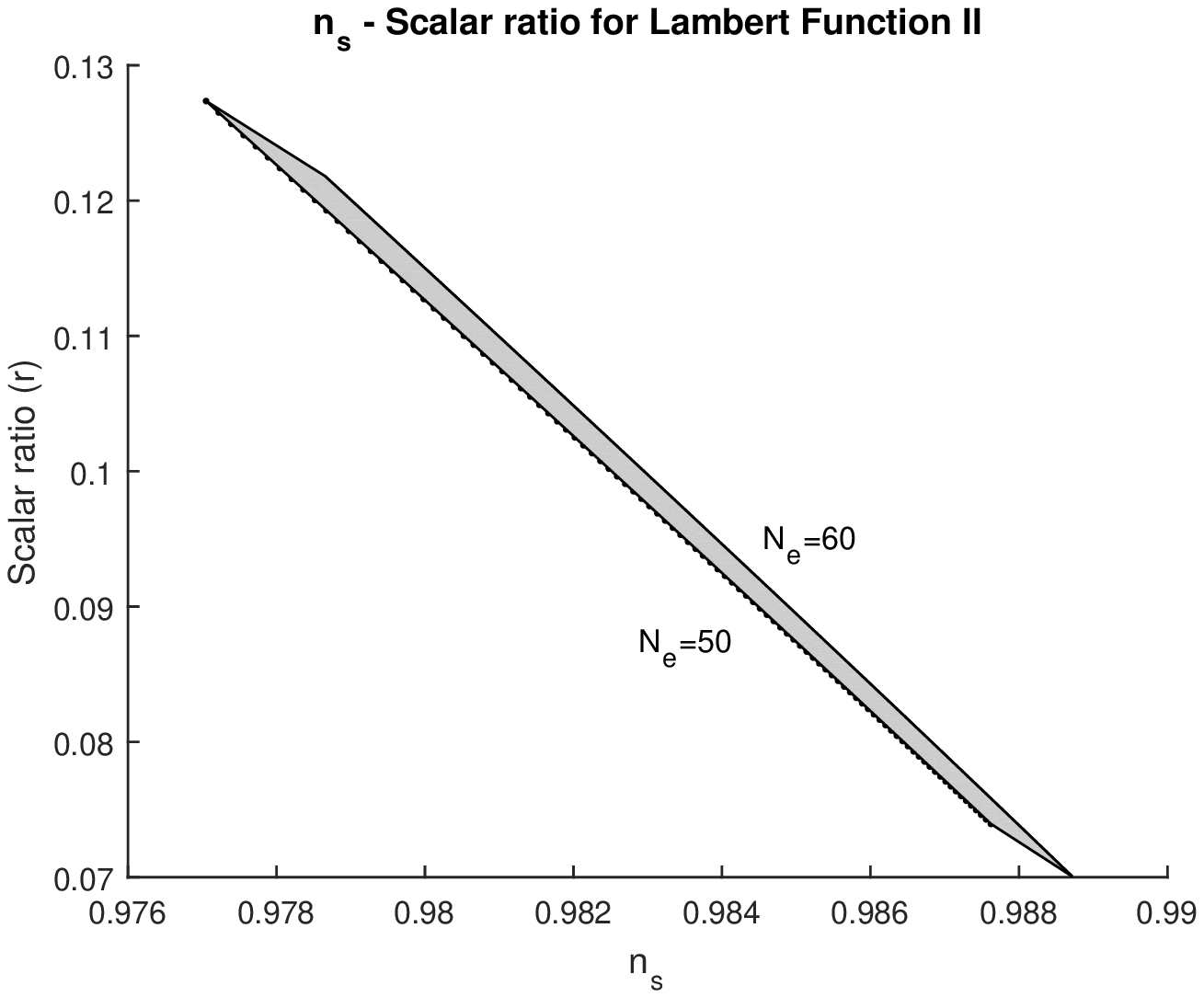}
\label{fig3a} \centering
\includegraphics[scale=0.55]{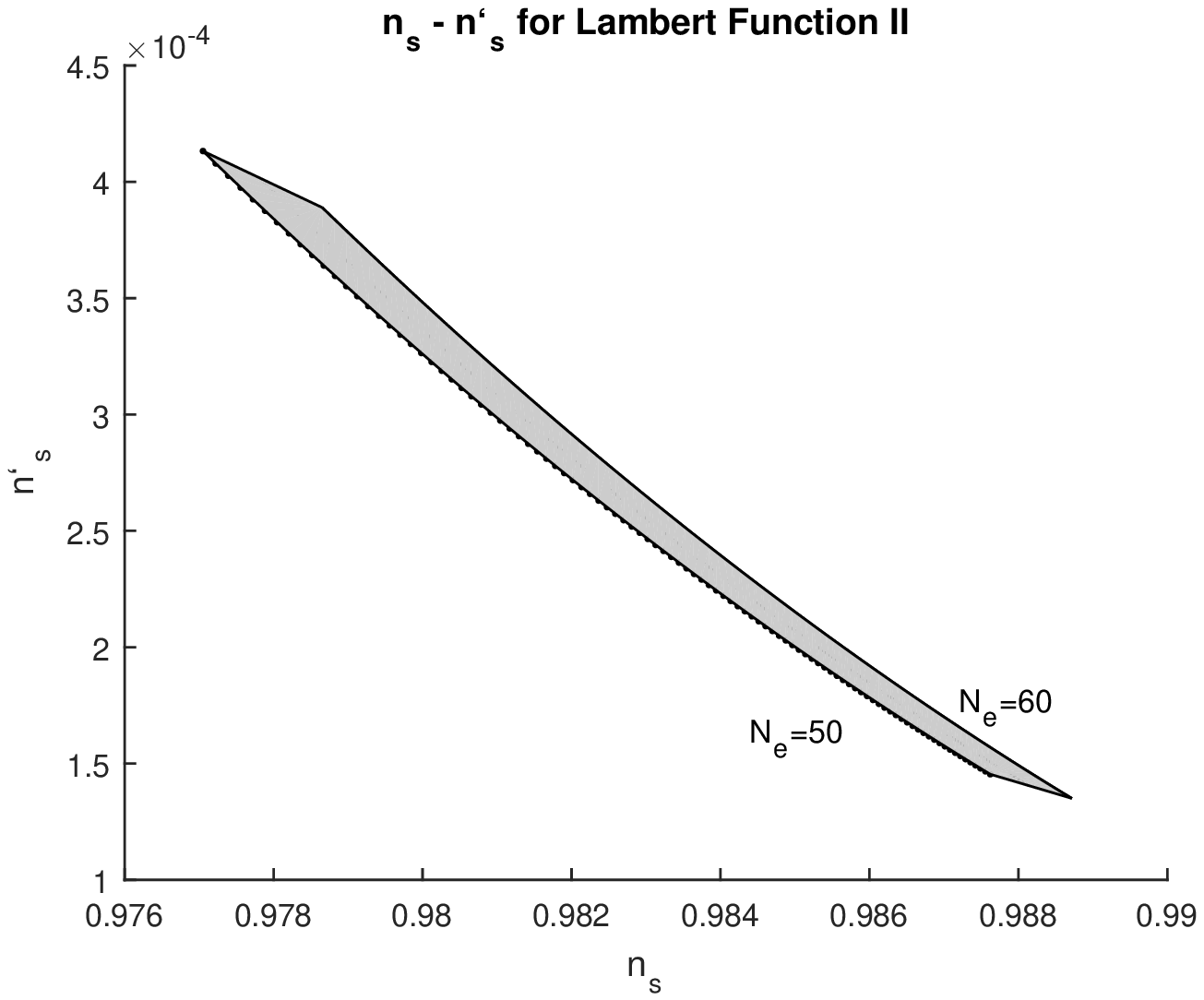}
\label{fig1b} \caption{The left figure is the $n_{s}-r$ \ diagram for the
model \textquotedblleft Lambert function II\textquotedblright. The right
figure is $\ $the $n_{s}-n_{s}^{\prime}$ diagram for the same model. The plots
are for $N_{e}\in\left[  50,60\right]  $ and $q\in\left[  65,100\right]  $.
$~$}%
\label{fig3}%
\end{figure}

In Fig \ref{fig3} the $n_{s}-r$ and $n_{s}-n_{s}^{\prime}$ diagrams \ are
given for $q\in\left[  65,100\right]  $, where we observe that $n_{s}^{\prime
}\rightarrow0$ as $n_{s}\rightarrow1$, while the relation $r\left(
n_{s}\right)  $ is linear. However, for $N_{e}\in\left[  50,60\right]  $ we
see that in the range of $n_{s}$ given by the Planck data we need to have
$r>0.11$. 

\subsection{Error function solution}

\begin{figure}[t]
\centering
\includegraphics[scale=0.55]{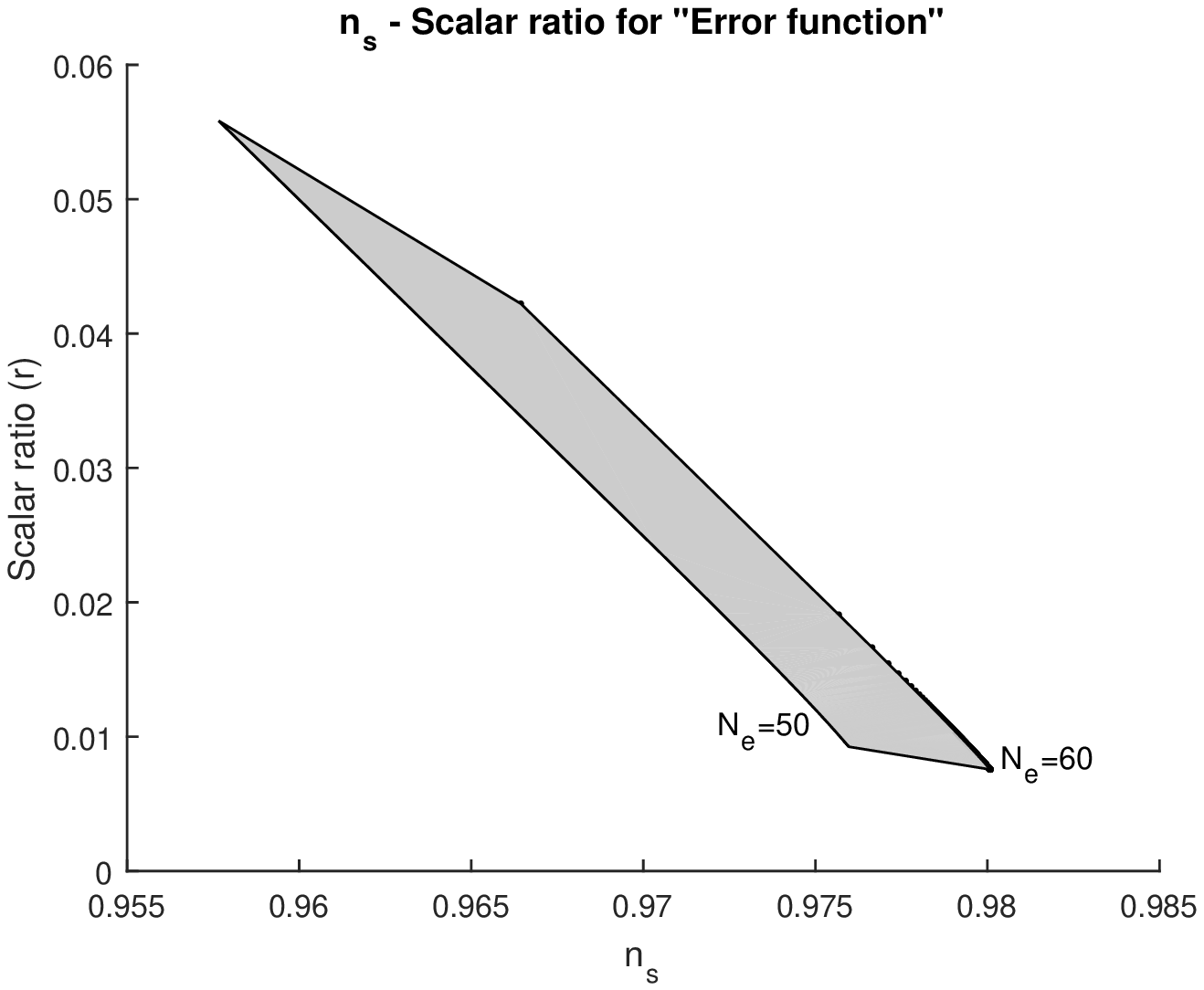}
\centering
\includegraphics[scale=0.55]{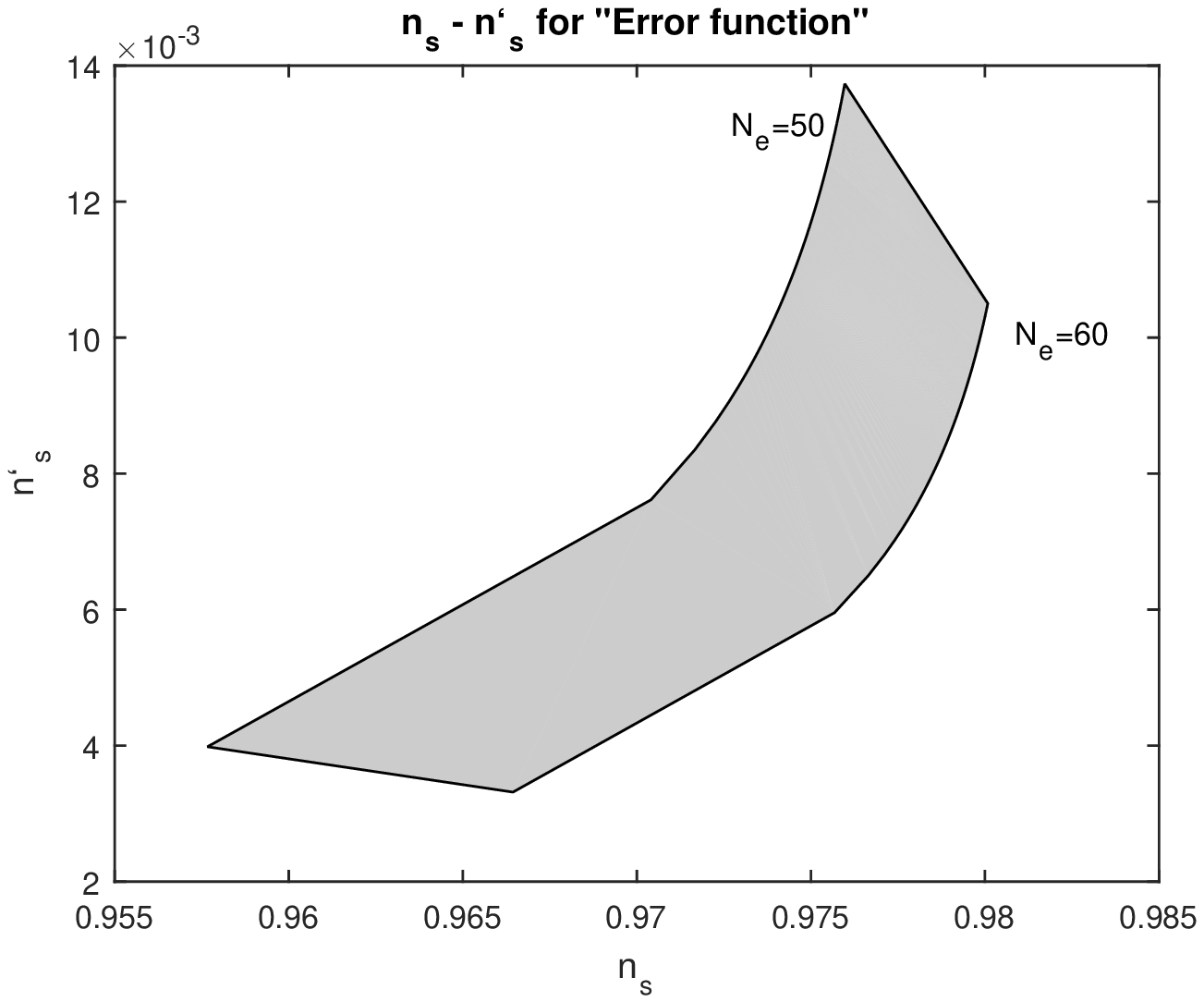}
\caption{The left figure is the $n_{s}-r$ \ diagram for the model
\textquotedblleft Error Function\textquotedblright. The right figure is
$\ $the $n_{s}-n_{s}^{\prime}$ diagram for the same model. The plots are for
$N_{e}\in\left[  50,60\right]  $ and $AB\in\left[  -100,-0.02\right]  $. $~$}%
\label{fig5}%
\end{figure}

For the \textquotedblleft Error function\textquotedblright\ solution for the
equation of state \ref{sd.48}, the HSR parameters are given by%
\begin{equation}
\varepsilon_{H}=-\frac{3}{\omega\ln\left(  AB\omega\right)  }~,~\eta_{H}%
=\frac{3}{\omega}~,~\xi_{H}=\frac{3}{2\sqrt{2}}\sqrt{\varepsilon_{H}}\eta
_{H}\left(  2\ln\left(  AB\omega\right)  -1\right)  ,
\end{equation}
and so inflation ends when $\omega$ reaches the value%
\begin{equation}
\omega_{f}=-\frac{3}{W\left(  -3AB\right)  }%
\end{equation}
where $W$ is the Lambert function. We can easily see that in order for
$\omega_{f}$ to be real,~$AB<0$. \ The $n_{s}-r$ and $n_{s}-n_{s}^{\prime}$
diagrams for this model are given in Fig.\ref{fig5} for the range of e-folds
$N_{e}\in\left[  50,60\right]  $ and for $AB\in\left[  -100,-0.02\right]  $.
From the plots we observe that for $r<0.06$, we have $n_{s}\in\left[
0.957,0.981\right]  $ and $n_{s}^{\prime}\in\left[  3,10\right]  \times
10^{-3}$. These values are compared with the best-fit values from the Planck
collaboration. 

\section{Conclusions}

\label{conc}

In this work we studied exact solutions in scalar field cosmology using a new
mathematical approach, and with emphasis on inflationary models. We have found
new closed-form solutions for spatially flat FLRW universes with or without an
extra matter source. For the latter cosmological scenario, we determined exact
solutions for the case in which the scalar field mimics the perfect fluid, the
scalar field has a constant equation of state parameter different from that of
the perfect fluid, and when the scalar field provides two perfect-fluid terms
in the field equations. The first solution is the well known special solution
of the exponential potential, while in the other two solutions the scalar
field potentials are expressed in hyper trigonometric functions and the
unified cold dark matter potential is recovered. \ Furthermore, these
expressions can be applied in order to construct other solutions in a FLRW
spacetime with spatial curvature.

In the cosmological scenario in which the universe is dominated by the scalar
field we determined the scalar field model in which the equivalent equation of
state parameter is that of the Chaplygin gas, or some generalizations of the
Chaplygin gas which have been proposed in the literature. \ We also considered
solutions in which the Hubble function is expressed in terms of the Lambert
function or by logarithmic function. These models provide exact inflationary
universe solutions.

We compared these solutions with the constraints on inflation from the Planck
2015 collaboration. In order to perform this analysis we expressed the Hubble
slow-roll (HSR) parameters in terms of the expansion scale factor in the
variables defined by our solution-generating functions. For every specific
model and solution we calculated the HSR parameters and we derived the
spectral indices in the first approximation. The diagrams for the density
perturbations $\left(  n_{s}\right)  ~$with the scalar ratio $\left(
r\right)  $ and the variation $n_{s}^{\prime}$ have been derived and the
subset of models which are compatible with the Planck 2015 data set are delineated.

\begin{acknowledgments}
JDB acknowledges support from the STFC. AP acknowledges financial support of
FONDECYT grant no. 3160121.
\end{acknowledgments}

\end{document}